\documentclass[
 pre,
 amsmath,amssymb,
preprint,%
]{revtex4-2}

\usepackage{graphicx}
\usepackage{dcolumn}
\usepackage{bm}
\usepackage{subcaption}
\usepackage{verbatim}
\begin{document}

\preprint{}

\title{Activity and Competing Length Scales in an Anomalous Core-Softened Fluid}

\author{Davi Felipe Kray Silva}
 \affiliation{Programa de Pós-Graduação em Física, Instituto de Física e Matemática, Universidade Federal de Pelotas, Pelotas, RS, Brazil}

\author{Thiago Puccinelli}
 \affiliation{Departamento de Física, Centro de Ciências Exatas e Naturais, Universidade Federal de Santa Maria, Santa Maria, RS, Brazil}

\author{Walas Silva-Oliveira}
 \affiliation{Programa de Pós-Graduação em Modelagem Matemática, Instituto de Física e Matemática, Universidade Federal de Pelotas. Caixa Postal 354, CEP 96001-970, Pelotas, RS, Brazil}
 
\author{Leandro B. Krott}%
 \affiliation{Centro de Ciências, Tecnologias e Saúde, Campus Araranguá, Universidade Federal de Santa Catarina, Araranguá, SC, Brazil.}%

\author{José Rafael Bordin}
\affiliation{Fachbereich Physik, Universität Konstanz, Konstanz, Deutschland}
 \affiliation{Departamento de Física, Instituto de Física e Matemática, Universidade Federal de Pelotas, Pelotas, RS, Brazil}
\email{jrbordin@ufpel.edu.br}
\date{\today}

\begin{abstract}
The interplay between activity and competing interaction length scales remains largely unexplored, despite its relevance to many soft and biological systems. Here, we study Active Brownian Particles interacting through a ramp-like core-softened potential that exhibits water-like anomalies in equilibrium. By varying the activity over a broad range of densities along two representative isotherms, one within the anomalous region and the other above it, we examine how self-propulsion modifies the structure and dynamics of the fluid. To gain microscopic insight into these changes, we construct effective interactions from the steady-state pair correlations using iterative Boltzmann inversion. We find that activity progressively suppresses the anomalies of the passive fluid, although signatures of the underlying structural crossover remain visible in normalized quantities. The effective interactions reveal that self-propulsion lowers the distinction between the local environments and facilitates population transfer between the two characteristic length scales. These results indicate that activity primarily acts by facilitating population transfer between the two local environments, thereby reducing the structural competition responsible for the anomalous response.\end{abstract}

\maketitle

\section{Introduction}

Active matter systems provide a natural framework to investigate how nonequilibrium driving modifies the structural and dynamical properties of interacting particle systems~\cite{Ramaswamy2010,DeMagistris2015, Liebchen2021}. In particular, Active Brownian Particles (ABPs) have become a standard model to study how persistent self-propulsion competes with interparticle interactions and thermal fluctuations~\cite{Romanczuk2012,Bechinger16}. This competition leads to a variety of collective phenomena, including motility-induced phase separation, clustering, and nontrivial transport properties~\cite{Hallatschek2023,teVrugt2025}. However, most studies have focused on systems with simple, single-length-scale interactions, where activity mainly competes with thermal motion and effective attractions~\cite{Gompper2025}.

In contrast, many complex fluids of physical, biological, and technological relevance are governed by interactions with more than one characteristic length scale~\cite{Bishop2023,Zhang2017,SMarques2020}. Examples include soft colloids, microgels, macromolecular assemblies, and biological cells, where the interplay between excluded volume, deformability, and longer-range interactions can generate competing local structures and multiple preferred interparticle distances.~\cite{Likos2001, Likos2006,Alvarado2026,Aranson2022,Deblais2023,Manning2023,Fu2022,Hopkins2023}. In equilibrium, the competition between these length scales is known to generate anomalous behaviors, such as non-monotonic diffusion and structural rearrangements~\cite{Bordin2023,Cardoso2021}. Core-softened potentials, particularly ramp-like interactions, provide a simple model to capture this physics, allowing one to isolate the role of competing length scales without additional complexity~\cite{Jagla1999,franzese2011,hemmer70,roca22}.

While core-softened models are well understood in equilibrium, their behavior under nonequilibrium driving remains less explored. In active systems, self-propulsion can modify interparticle correlations and even induce emergent length scales in the steady state~\cite{ReesZimmerman2026,Farage2015,Sarkar2025}. In this sense, core-softened models provide a minimal framework to investigate how activity acts on systems with competing local structures.

Recent studies have started to address ABPs interacting through two-length-scale potentials, showing that activity can strongly affect structural organization and phase behavior. In particular, self-propulsion may help particles overcome the energetic barrier associated with the outer shell, favoring configurations at shorter distances~\cite{roca22}. Likewise, active perturbations in anomalous liquids have been shown to enhance mobility and fluidize structured states~\cite{Truong2025}. However, the impact of activity on the anomalous properties themselves, especially those arising from the competition between characteristic length scales, remains poorly understood.

In particular, it remains unclear whether activity merely changes the populations associated with the competing length scales or whether it fundamentally alters the microscopic mechanism responsible for water-like anomalies. Since these anomalies originate from the redistribution of particles between distinct local environments~\cite{Shi2018,Russo2014}, understanding how self-propulsion modifies their energetic accessibility is essential not only for connecting nonequilibrium driving with anomalous behavior, but also for clarifying whether the same microscopic mechanisms may influence the collective organization of active fluids.

In this work, we study ABPs interacting through a ramp-like core-softened potential that exhibits water-like anomalies in both two and three dimensions~\cite{Oliveira06a,deOliveira2006,Bordin2018}. By varying the activity over a broad range of densities and considering both anomalous and non-anomalous temperature regimes, we investigate how self-propulsion modifies the structural and dynamical properties of a fluid with competing interaction length scales. To gain microscopic insight into these changes, we construct an effective interaction from the steady-state pair correlations using iterative Boltzmann inversion. This approach allows us to directly examine how activity modifies the energetic accessibility of the competing local environments and the consequences of these changes for the structural, dynamical, and nonequilibrium behavior of the fluid.

The remainder of this paper is organized as follows. In Sec.~II, we introduce the model and simulation details. Section~III presents the structural, dynamical, and effective-interaction results. Finally, Sec.~IV summarizes our main conclusions and perspectives.

\section{The Model and Simulation Details}

We consider a two–dimensional system of monomeric ABPs interacting via an isotropic core–softened (CS) potential. Particles interact via a two–length–scale core–softened potential given by a Lennard–Jones term plus a Gaussian shoulder~\cite{Oliveira06a,Bordin2023}:
\begin{equation}
u(r) = 4\varepsilon\left[\left(\frac{\sigma}{r}\right)^{12} - \left(\frac{\sigma}{r}\right)^6\right] 
+ u_0 \exp\!\left[-\frac{1}{c_0^2}\left(\frac{r - r_0}{\sigma}\right)^2\right].
\end{equation}
We fix $r_0/\sigma = 0.7$, $u_0 = 5\varepsilon$, and $c_0^2 = 1.0$, yielding two preferred distances at $r \approx 1.2\sigma$ and $r \approx 2.0\sigma$~\cite{BarrosdeOliveira2010}. A cutoff $r_c = 3.5\sigma$ was used.

Simulations were performed using the LAMMPS molecular dynamics package~\cite{lammps_citation}, employing reduced Lennard–Jones (LJ) units, where length, energy, and mass are given in units of $\sigma$, $\varepsilon$, and $m$, respectively. This setup allows us to directly probe how activity modifies the competition between the two characteristic length scales that govern the anomalous behavior of the passive system. Previous studies have shown that this potential exhibits water-like anomalies in both two and three dimensions~\cite{Oliveira06a, deOliveira2006,Bordin2018, Bordin2023, Cardoso2021}. The present two-dimensional system was selected as a minimal framework to isolate the interplay between activity and competing length scales.

The system consists of $N = 10000$ particles in a square box of area $A = N/\rho$, with periodic boundary conditions in both directions. We performed Langevin dynamics simulations at $T = 0.2$ (anomalous) and $T = 1.0$ (non-anomalous). These temperatures were chosen to represent two limiting situations of the passive system: one where competition between local environments produces well-defined anomalies and another where only remnants of the structural crossover remain~\cite{Cardoso2021}. Densities span the range $\rho = 0.01$ to $0.55$, covering dilute to dense states.

Particle dynamics follow the overdamped Active Brownian Particle model implemented through the Brownian dynamics integrator (fix brownian) available in LAMMPS~\cite{lammps_citation}, where each particle self–propels with speed $v_p$ along an orientation undergoing rotational diffusion. We set $\gamma_t = 15.0$ and $\gamma_r = 0.1$. The strength of activity can be characterized by the Peclet number,
\begin{equation}
Pe = \frac{v_p}{D_r \sigma},
\end{equation}
where $D_r$ is the rotational diffusion coefficient, set by the parameters of the Brownian integrator and determines the orientational persistence of the active particles. In this work, we vary $Pe$ from $0.0$, corresponding to the passive case, up to $Pe = 5.0$.

Thermal noise is introduced via Langevin terms consistent with the fluctuation–dissipation theorem. The equations of motion are integrated with timestep $\delta t = 0.001$. Configurations are initialized at low density and equilibrated at the target state point for $10^6$ steps, followed by $10^7$ production steps. Steady state is verified from the time evolution of thermodynamic and structural observables. Statistical uncertainties were estimated by block averaging over independent time windows.

Dynamics are characterized through the mean–squared displacement (MSD),
\begin{equation}
\langle \delta r^2(t) \rangle = \left\langle \left| \mathbf{r}(t) - \mathbf{r}(0) \right|^2 \right\rangle.
\end{equation}
At sufficiently long times, the system approaches a diffusive regime for all explored conditions. The diffusion coefficient is then obtained using the Einstein relation in two dimensions,
\begin{equation}
D = \lim_{t\to\infty}\frac{\langle \delta r^2(t)\rangle}{4t}.
\end{equation}

Structural properties are analyzed via the radial distribution function (RDF) $g(r)$ and the two–body excess entropy $s_2$~\cite{raveche1971,Baranyai1989},
\begin{equation}
s_2 = -2\pi \rho \int_0^\infty \big[g(r)\ln g(r) - g(r) + 1\big] r \, dr,
\end{equation}
which provides a sensitive measure of translational order. The quantity $s_2$ is further used to quantify changes induced by activity relative to the passive system, and its time evolution is monitored to ensure convergence to steady state.

To gain microscopic insight into how activity modifies local structure, we constructed an effective pair interaction that reproduces the steady-state radial distribution function (RDF) of the active system~\cite{Farage2015,Alamarza04}. Since ABPs are intrinsically out of equilibrium, the resulting interaction should be interpreted as a state-dependent structural effective potential rather than as a true thermodynamic pair potential. Accordingly, the reconstructed interaction should be regarded as a structural mapping of the nonequilibrium steady state, rather than as a unique effective free-energy landscape. As an initial estimate, we employed the potential of mean force, $-\ln g_{\mathrm{target}}(r)$, where $\beta=(k_{\mathrm B}T)^{-1}$ and $g_{\mathrm{target}}(r)$ is the RDF measured in the active simulations. To account for finite-density effects, the interaction was subsequently refined using iterative Boltzmann inversion (IBI)~\cite{Reith2003, ReesZimmerman2026},

\begin{equation}
u_{\mathrm{eff}}^{(i+1)}(r)
=
u_{\mathrm{eff}}^{(i)}(r)
+
k_{\mathrm B}T
\ln
\left[
\frac{g^{(i)}(r)}
{g_{\mathrm{target}}(r)}
\right],
\end{equation}
where $g^{(i)}(r)$ is the RDF obtained from a passive simulation employing $u_{\mathrm{eff}}^{(i)}(r)$. The iterations were repeated using the ESPResSo package~\cite{Weik2019,Weeber2024} until the RDF of the equivalent passive system reproduced the target RDF of the active fluid. At very short separations, the RDF becomes poorly sampled due to the strong repulsive core of the original core-softened interaction. In this region, the effective interaction was smoothly extrapolated solely to ensure a numerically stable tabulated potential. The extrapolation was restricted to distances for which $g(r)\lesssim10^{-4}$, corresponding to configurations that occur with negligible probability in the simulations. Consequently, this regularization does not affect the resulting pair correlations or the structural mapping provided by the IBI procedure.

\section{Results and Discussions}

\subsection{Anomalous Region - $T = 0.20$}

\begin{figure}[h]
\centering
\begin{subfigure}[b]{0.49\textwidth}
        \includegraphics[width=\textwidth]{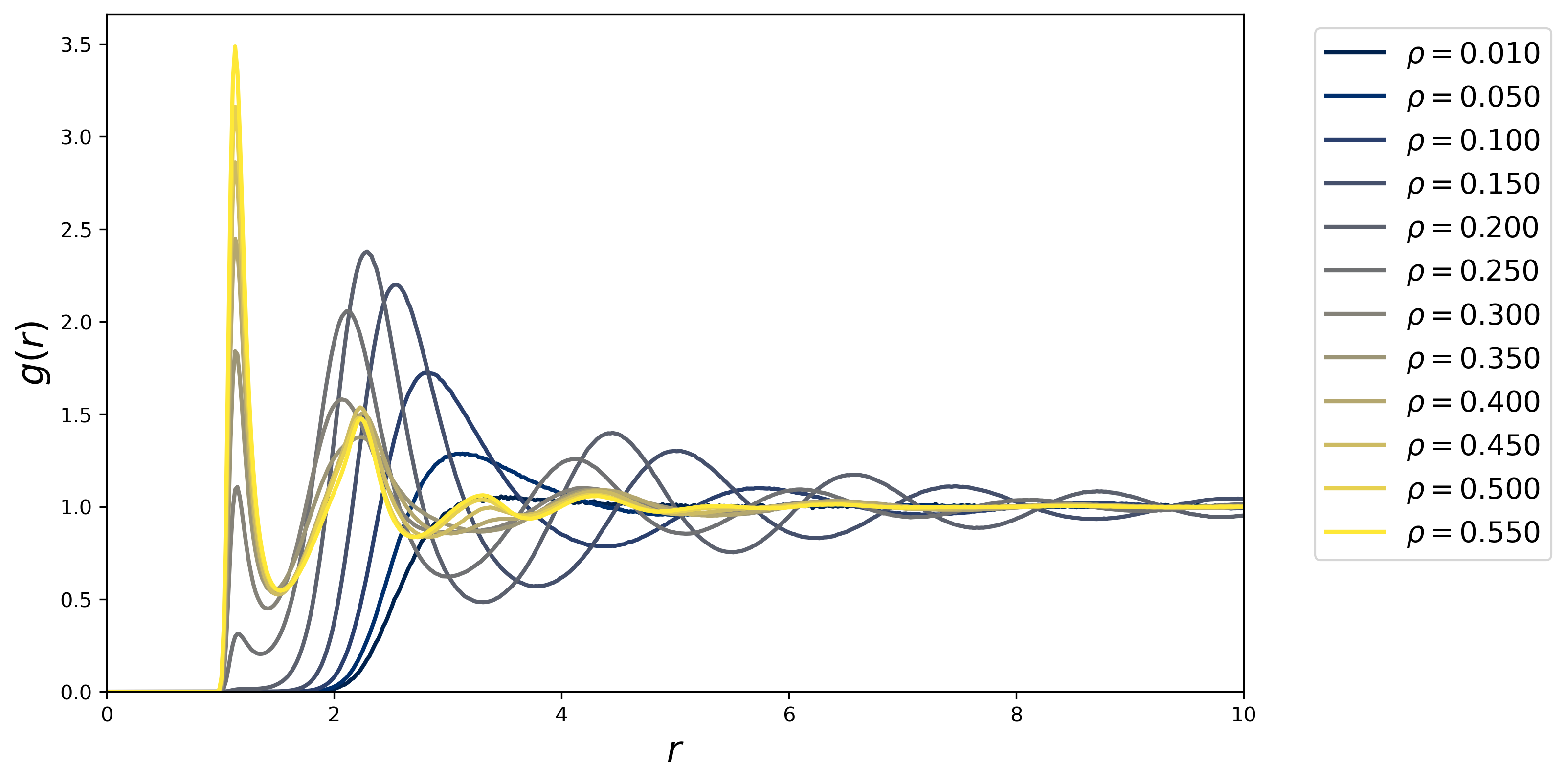}
\caption{}
    \end{subfigure}
\begin{subfigure}[b]{0.49\textwidth}
        \includegraphics[width=\textwidth]{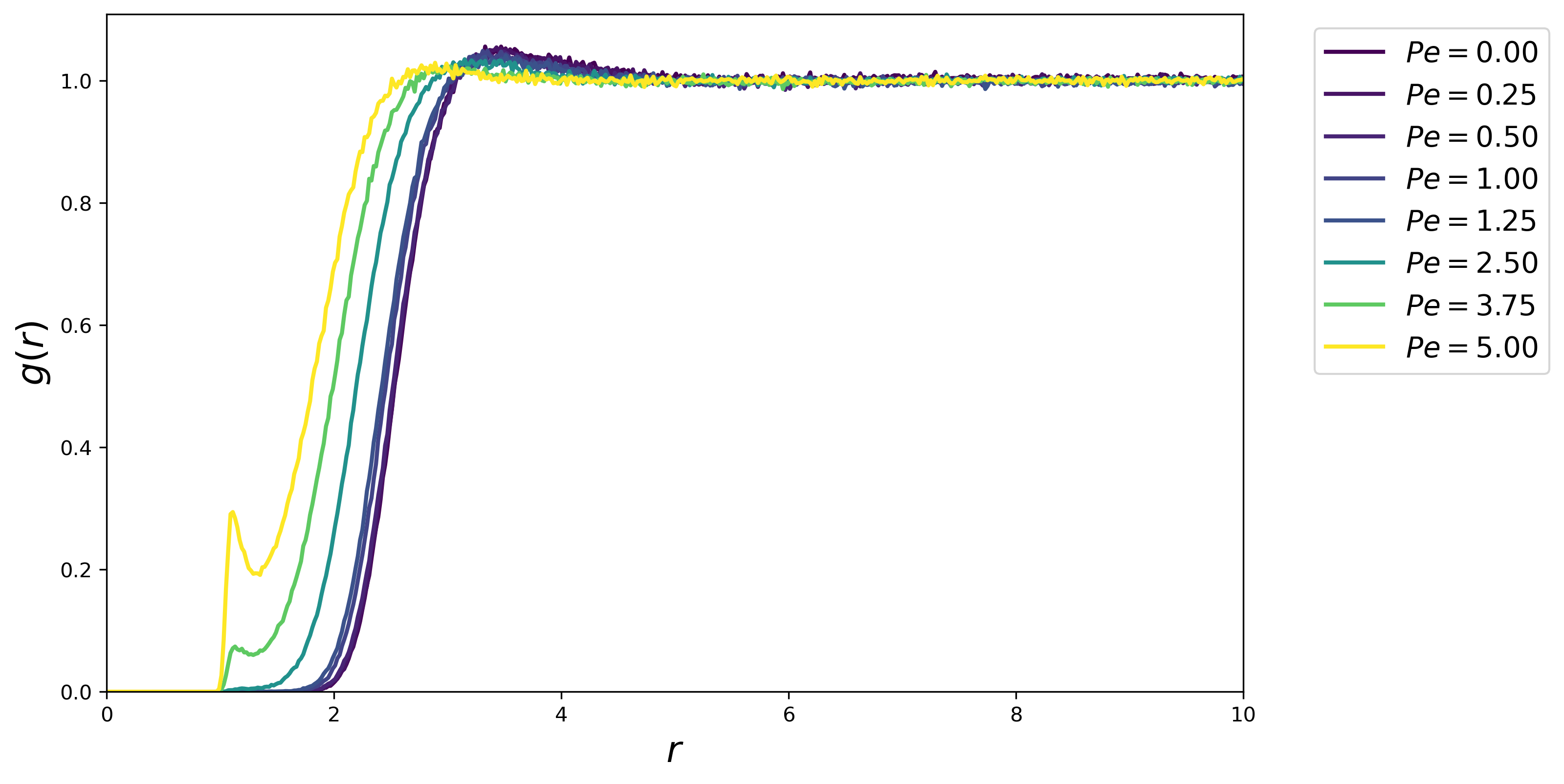}
\caption{}
    \end{subfigure}
    \begin{subfigure}[b]{0.49\textwidth}
        \includegraphics[width=\textwidth]{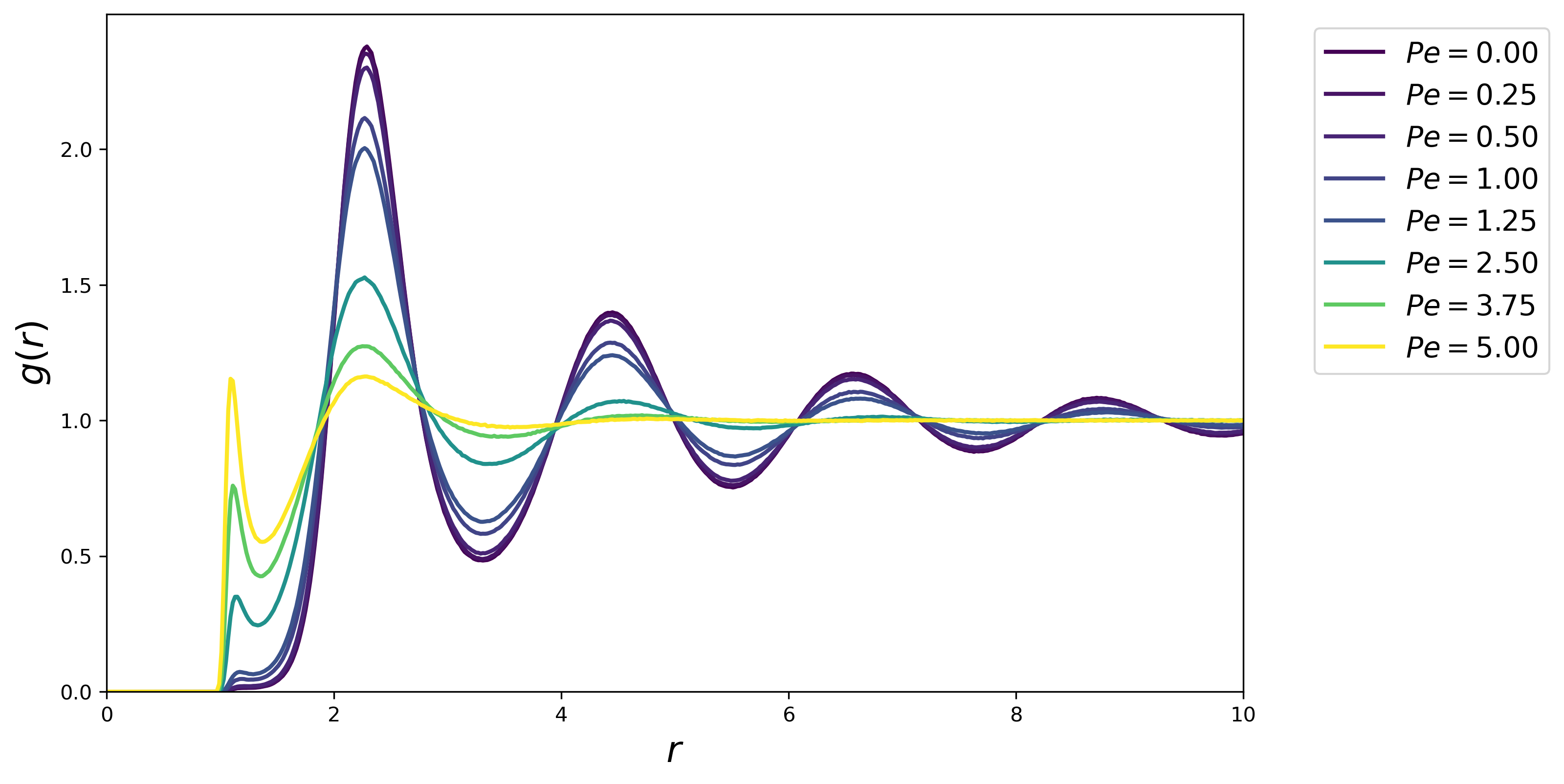}
\caption{}
    \end{subfigure}
    \begin{subfigure}[b]{0.49\textwidth}
        \includegraphics[width=\textwidth]{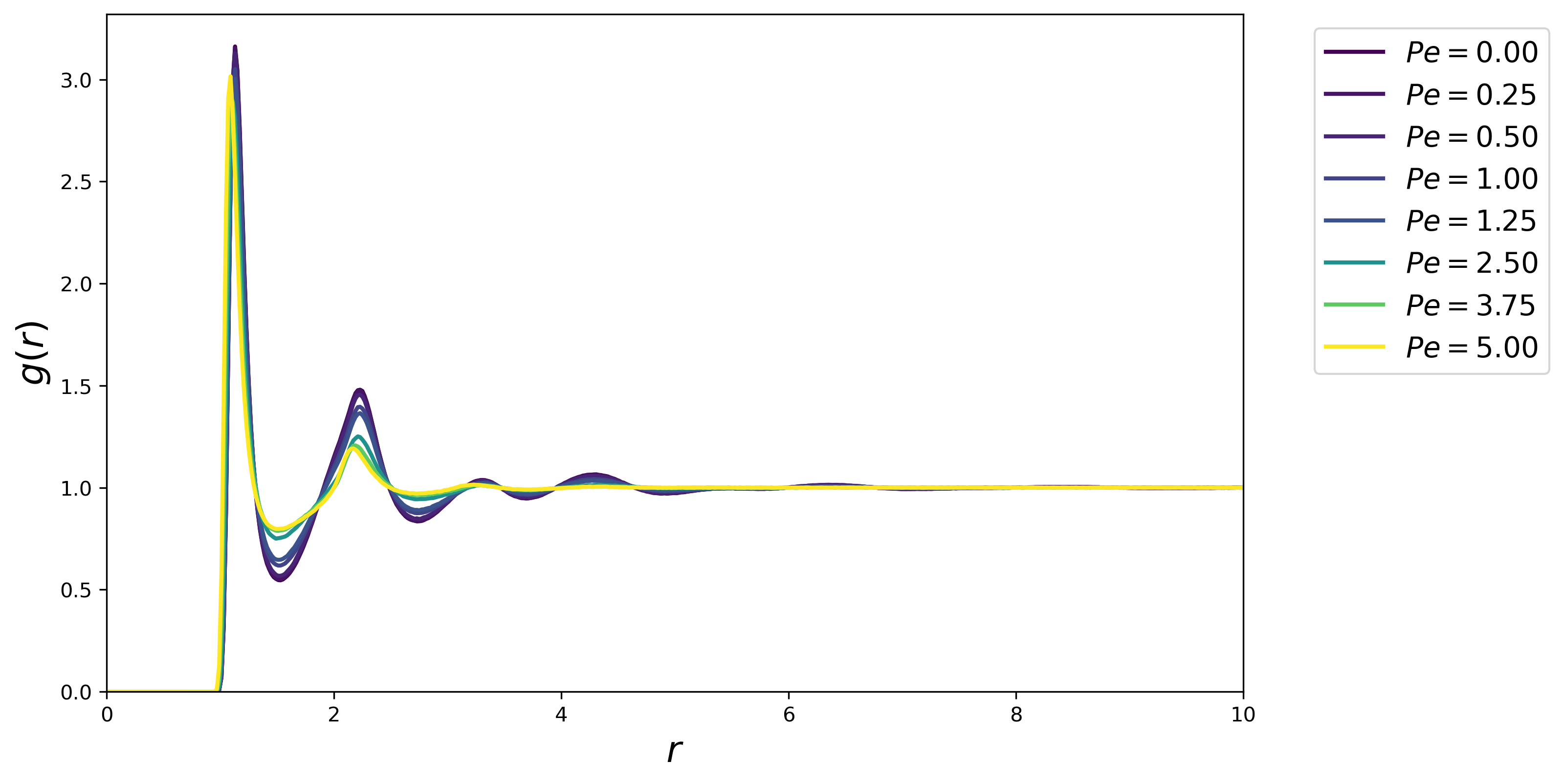}
\caption{}
    \end{subfigure}
\caption{(a) Radial distribution functions at $T=0.20$ for the passive system and different densities. Radial distribution functions for increasing activity at (b) $\rho=0.01$, (c) $\rho=0.20$, and (d) $\rho=0.50$.}
\label{rdfT020}

\end{figure}

 We begin by examining the effect of activity at $T=0.20$, which lies inside the anomalous region of the passive system~\cite{Cardoso2021}. Figure~\ref{rdfT020}(a) shows the radial distribution functions (RDFs) of the passive fluid for densities ranging from the dilute regime, $\rho=0.010$, to the dense-fluid regime, $\rho=0.550$.

At very low density, $\rho=0.010$, the system displays gas-like behavior with weak structural correlations. The first peak of $g(r)$ appears around the second characteristic distance, $r\approx2.25$, indicating that particles preferentially occupy the outer scale. As the density increases, this peak shifts toward shorter distances, reaching approximately $r\approx2.00$. Around $\rho\approx0.20$, a second peak develops near the first length scale, $r\approx1.00$, and becomes progressively more pronounced with increasing density, while the population associated with the outer scale decreases. This continuous redistribution of particles between the two preferred distances constitutes the microscopic origin of the water-like anomalies observed in core-softened systems.

To assess the effect of activity, we focus on three representative densities. At $\rho=0.010$ [Fig.~\ref{rdfT020}(b)], increasing $Pe$ promotes a shift toward shorter interparticle distances. The population associated with the outer scale decreases slightly, while a small but clear occupation emerges at the inner scale. This behavior resembles the effect of increasing density or temperature in the passive fluid~\cite{Cardoso2021}, indicating that activity favors local configurations associated with the first length scale even in the dilute regime.

At $\rho=0.20$ [Fig.~\ref{rdfT020}(c)], the effect of activity is more pronounced. The population at the first length scale increases, whereas the second one becomes less populated, indicating a redistribution between the two characteristic distances. In the passive system, this density corresponds to the onset of longer-range correlations and marks the region where the anomalies are most evident~\cite{Cardoso2021,Ilha2025,Puccinelli2025}. As activity increases, the oscillations of $g(r)$ beyond the first coordination shell become weaker, indicating that self-propulsion modifies both local arrangements and medium-range structural organization. 

At $\rho=0.50$ [Fig.~\ref{rdfT020}(d)], the passive system is already dominated by configurations associated with the first length scale. In this regime, activity produces only minor changes in the local structure and mainly suppresses correlations beyond the first neighbors, leading to a gradual damping of the oscillations in $g(r)$.

\begin{figure}[h]
\centering
\begin{subfigure}[b]{0.49\textwidth}
        \includegraphics[width=\textwidth]{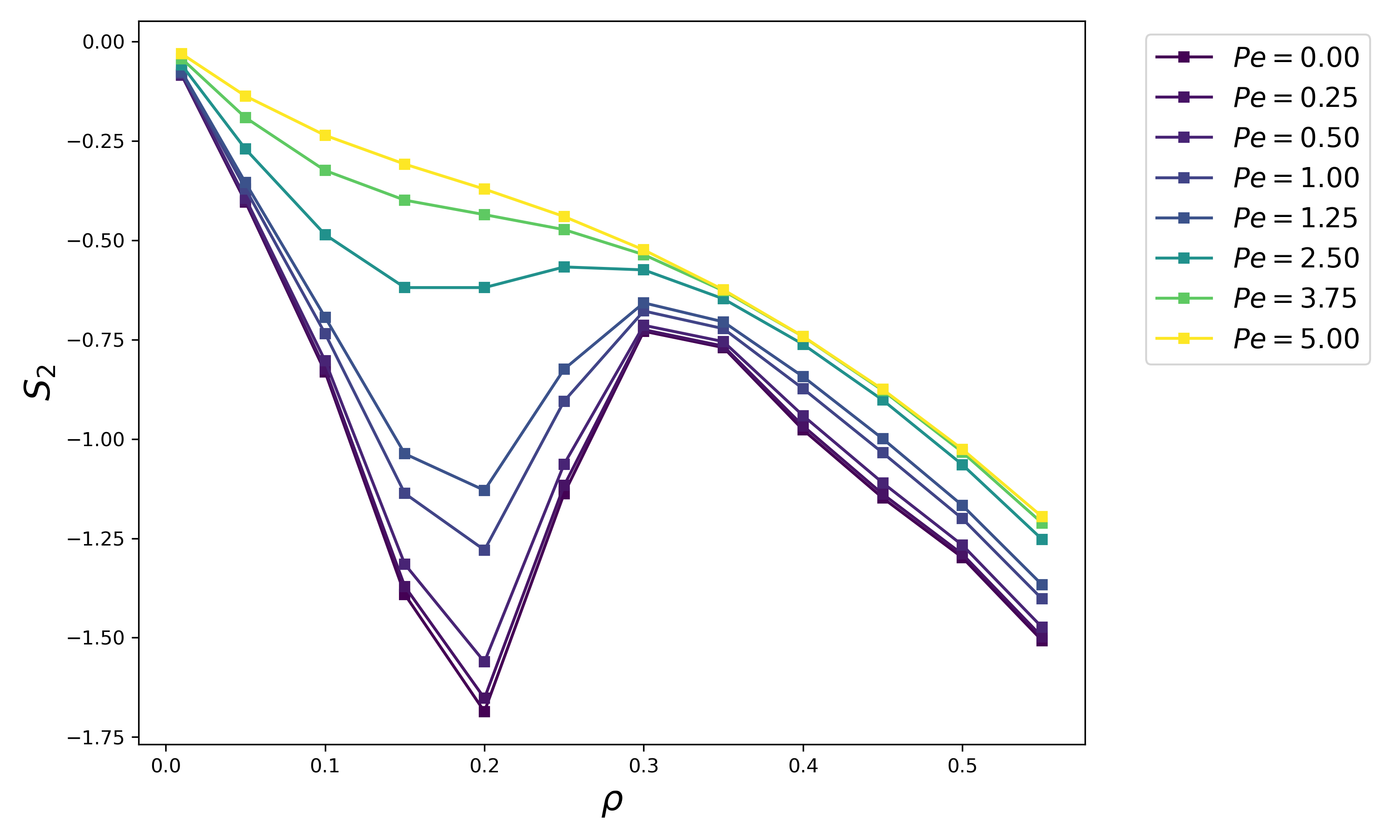}
\caption{}
    \end{subfigure}
\begin{subfigure}[b]{0.49\textwidth}
        \includegraphics[width=\textwidth]{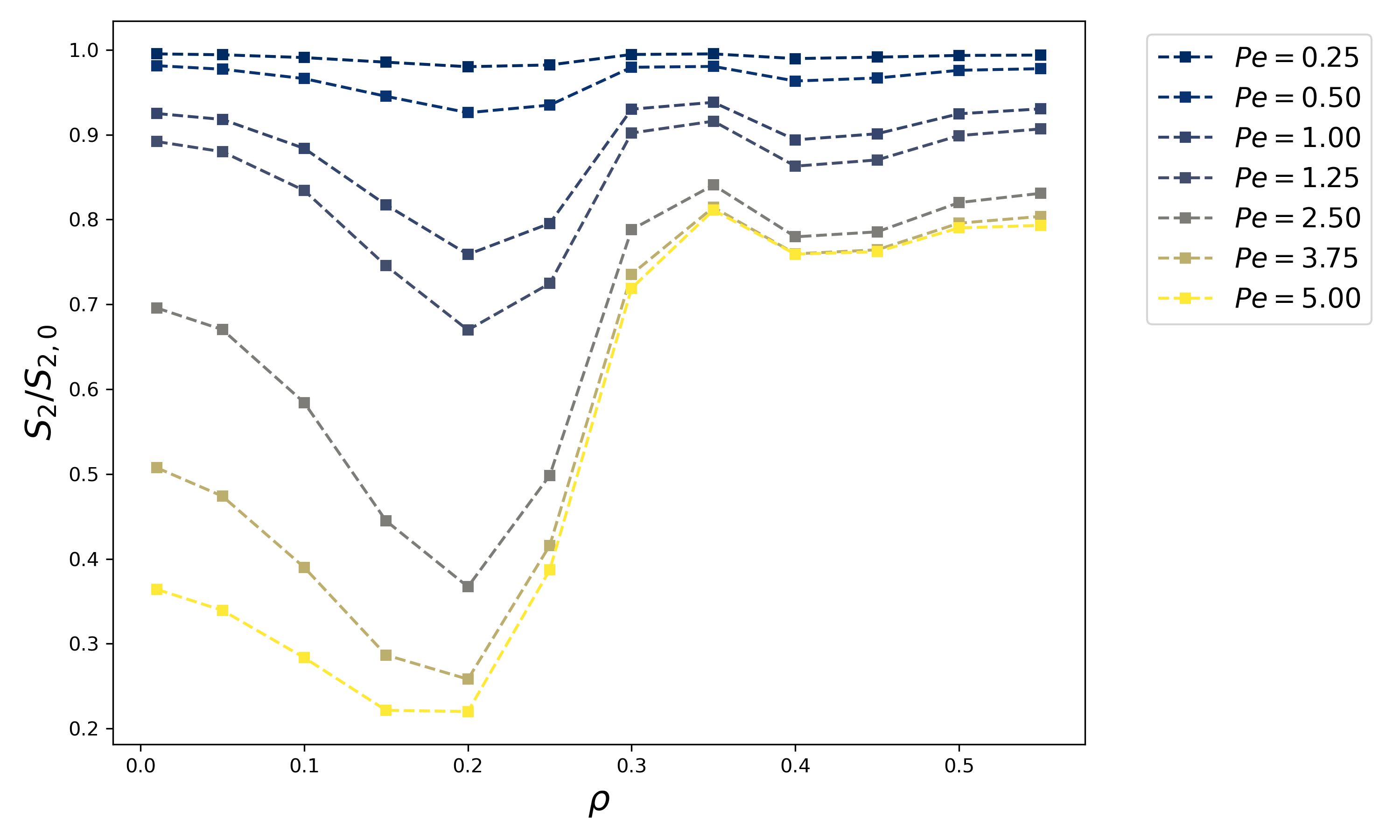}
\caption{}
    \end{subfigure}
\caption{(a) Pair excess entropy $S_2$ as a function of density for different activities at $T=0.20$. (b) Pair excess entropy normalized by its passive counterpart, $S_2/S_{2,0}$.}
\label{s2T020}
    \end{figure}

\begin{figure}[h]
\centering
\begin{subfigure}[b]{0.49\textwidth}
        \includegraphics[width=\textwidth]{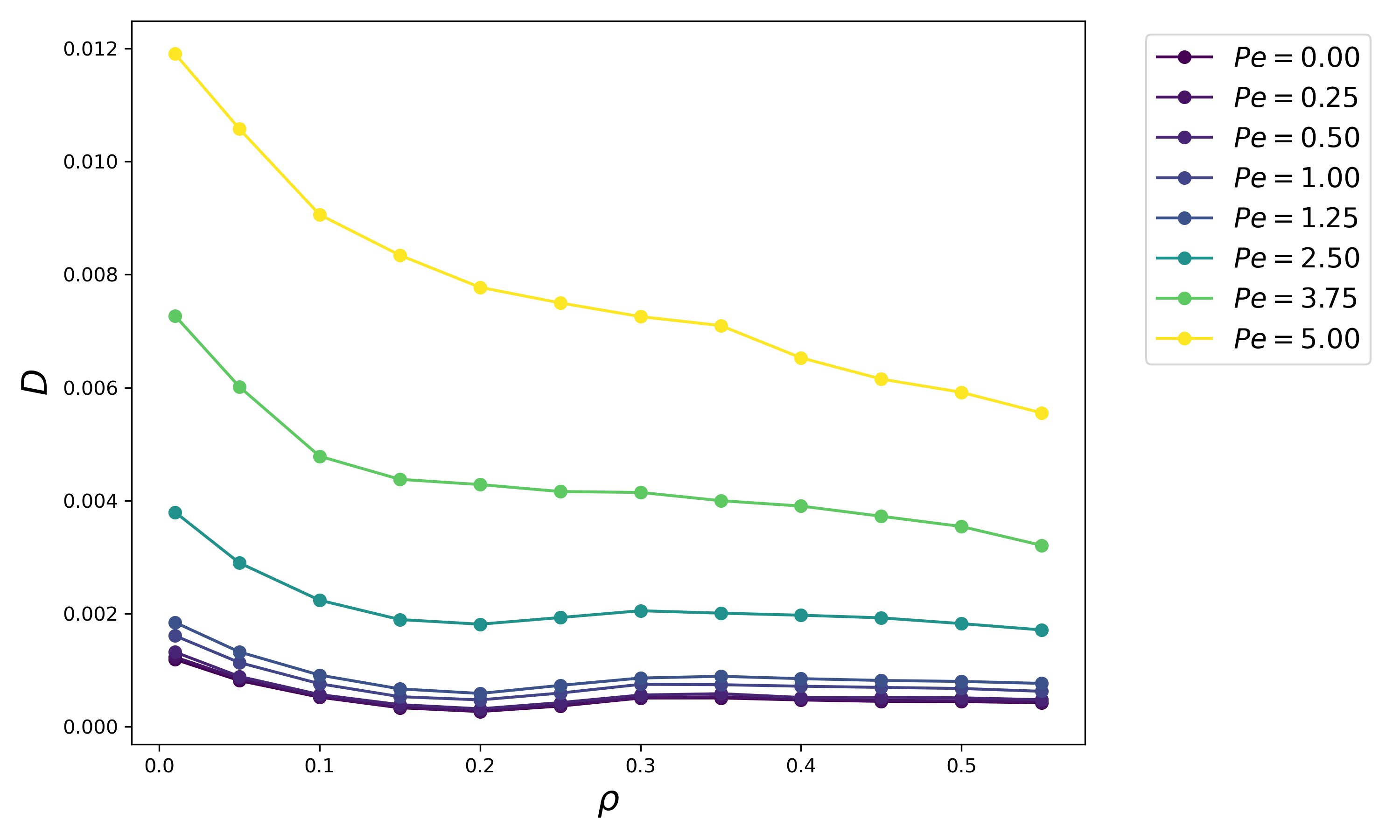}
\caption{}
    \end{subfigure}
\begin{subfigure}[b]{0.49\textwidth}
        \includegraphics[width=\textwidth]{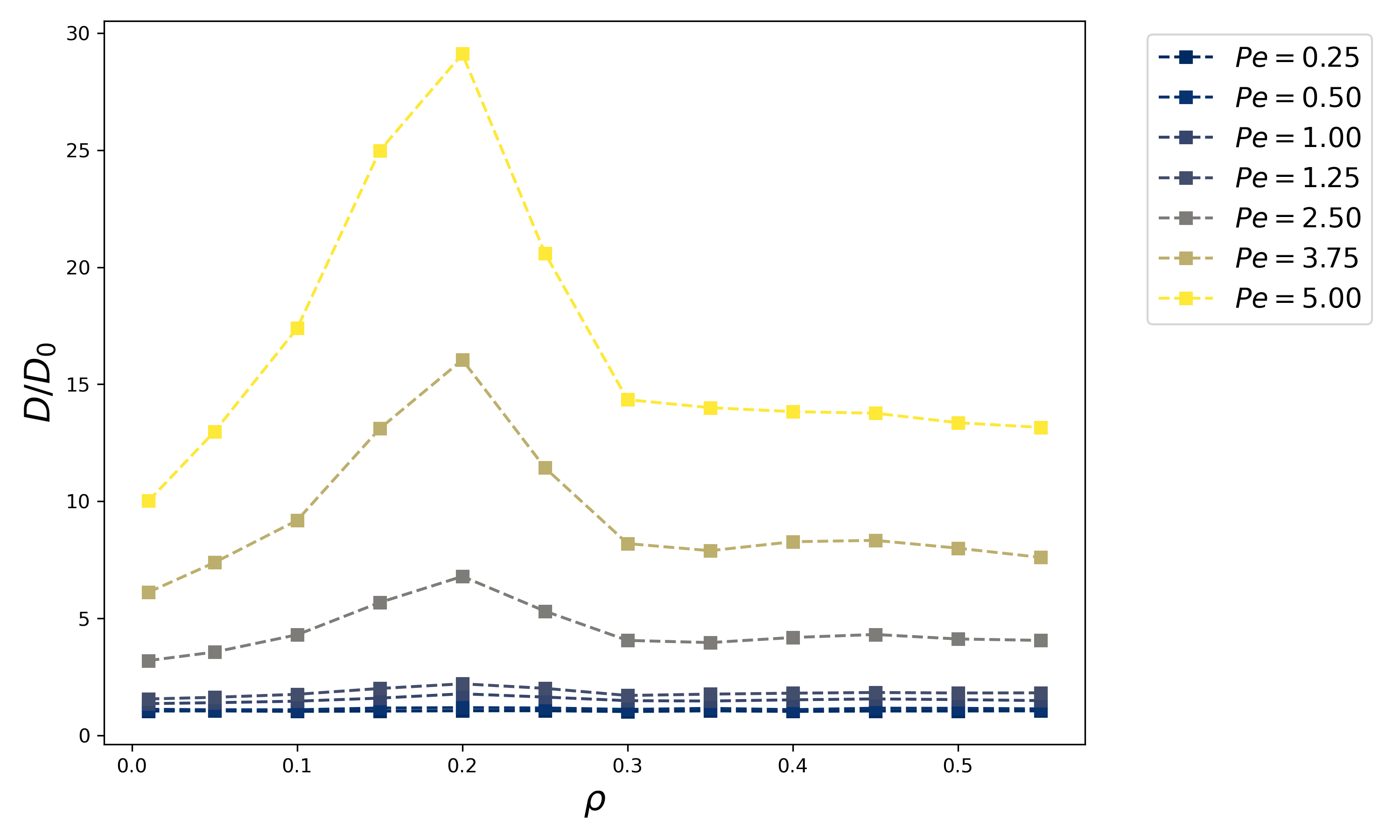}
\caption{}
    \end{subfigure}
    \caption{(a) Diffusion coefficient $D$ at $T = 0.20$ as function of density $\rho$ for distinct activities. (b) $D$ normalized by the passive diffusion, $D_{0}$, as function of $\rho$. }
    \label{DT020}

    \end{figure}

To quantify these structural changes, we analyze the pair excess entropy $S_2$ and the diffusion coefficient $D$ as functions of density. For the passive fluid [Figs.~\ref{s2T020}(a) and \ref{DT020}(a)], both quantities exhibit the expected anomalous behavior. The pair excess entropy presents a minimum around $\rho\approx0.20$ and a maximum near $\rho\approx0.30$, reflecting the redistribution of particles between the two characteristic distances. Over the same density range, the diffusion coefficient exhibits a non-monotonic dependence on density and increases as particles reorganize between local environments. The coincidence of these extrema indicates that structural rearrangements between the two length scales directly influence particle mobility.

As activity increases, these anomalies become less pronounced. The extrema in $S_2$ are attenuated and eventually disappear, while $D$ approaches a monotonic dependence on density. As shown below from the effective interactions, self-propulsion lowers the apparent barrier in the structural effective potential associated with occupying the inner length scale and promotes a redistribution of particle populations toward shorter separations. Consequently, the density range over which both local environments remain populated becomes narrower.

Remarkably, this suppression already occurs at relatively low Peclet numbers, indicating that the anomalous regime is highly sensitive to persistent nonequilibrium driving. Thus, the competition between the two local environments is not only weakened at strong activity, but begins to be disrupted as soon as self-propulsion becomes comparable to the structural relaxation associated with shell redistribution.

The underlying mechanism, however, remains visible in the normalized quantities. The ratio $S_2/S_{2,0}$ [Fig.~\ref{s2T020}(b)] retains a non-monotonic dependence on density, with extrema located approximately at the same densities as in the passive system. A similar trend is observed for $D/D_0$ [Fig.~\ref{DT020}(b)], which displays a maximum around $\rho\approx0.20$. Therefore, even though the anomalies disappear in absolute quantities, both structure and dynamics remain most sensitive to activity in the same density range where the competition between the two characteristic distances is strongest.

\begin{figure}[h]
\centering
    \begin{subfigure}[b]{0.49\textwidth}
        \includegraphics[width=\textwidth]{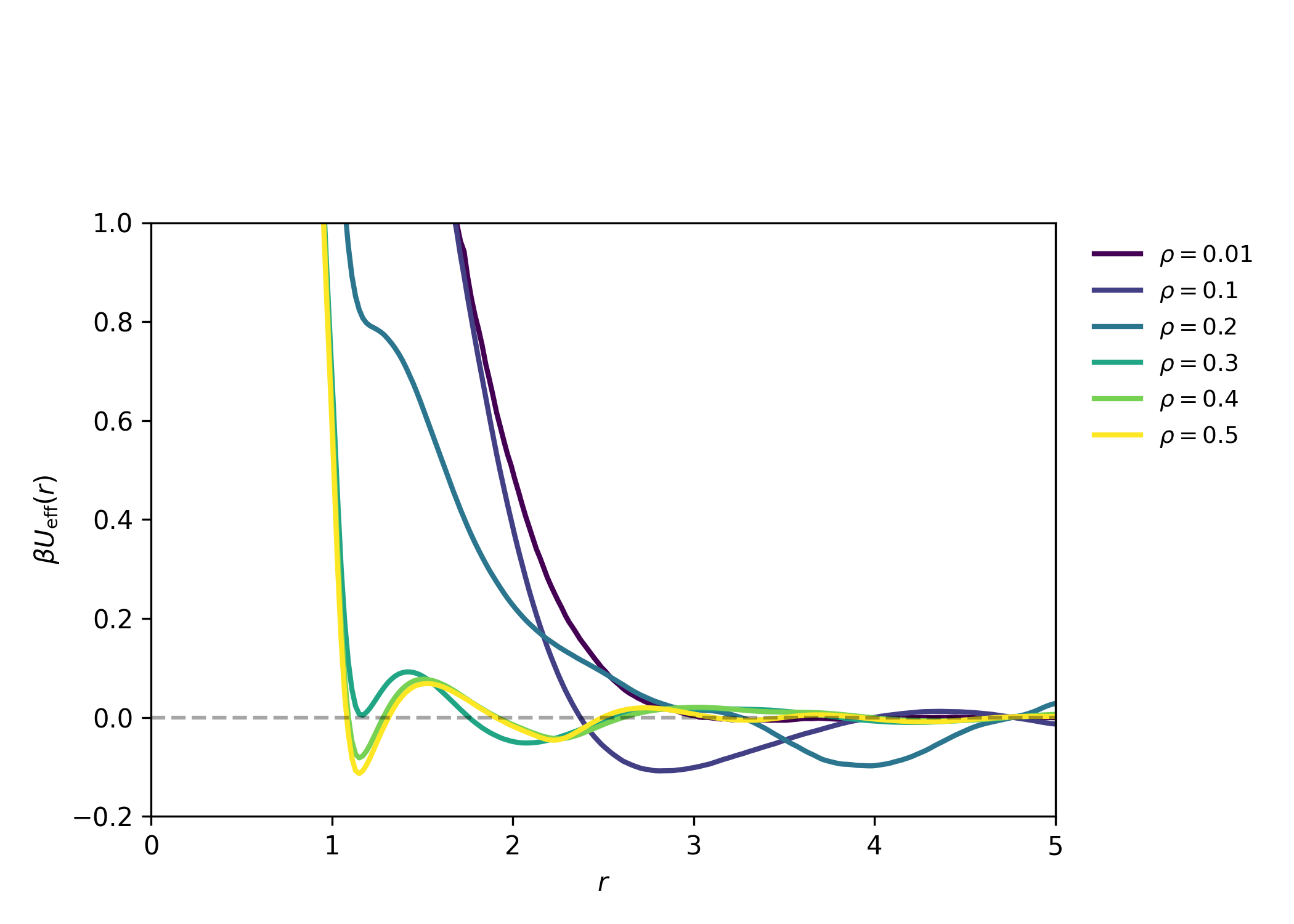}
\caption{$Pe = 0.0$}
    \end{subfigure}
\begin{subfigure}[b]{0.49\textwidth}
        \includegraphics[width=\textwidth]{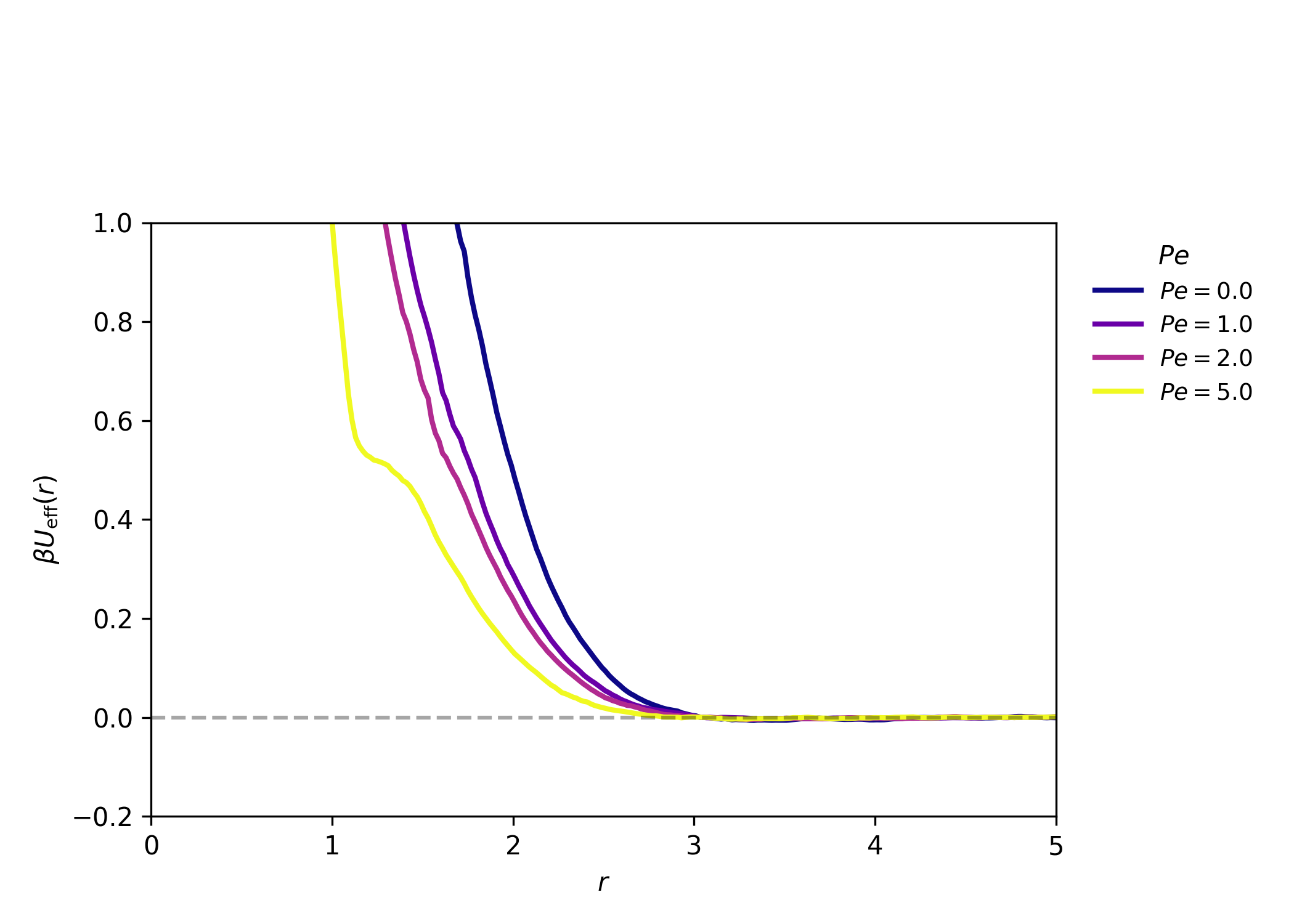}
\caption{$\rho = 0.01$}
    \end{subfigure}
\begin{subfigure}[b]{0.49\textwidth}
        \includegraphics[width=\textwidth]{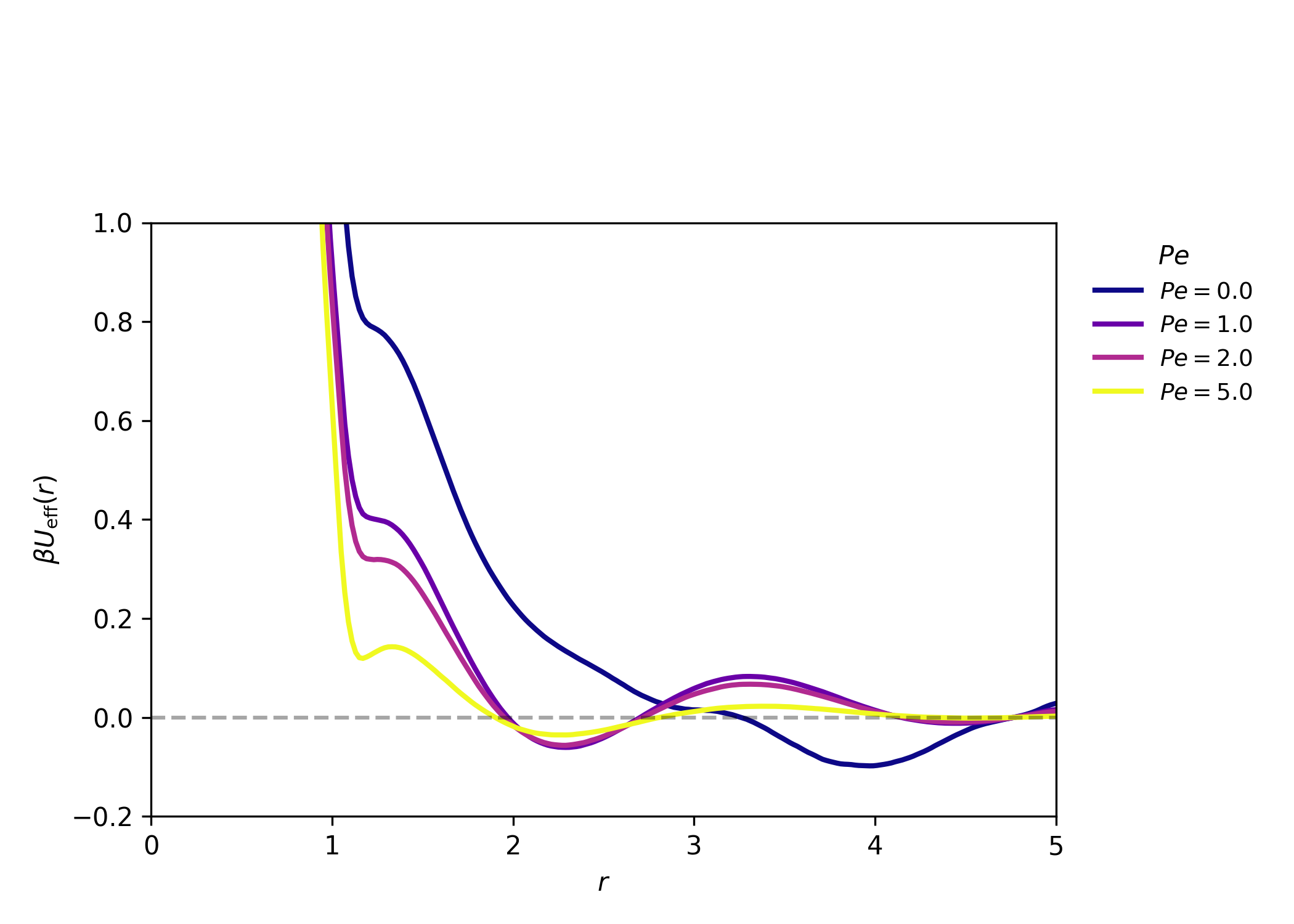}
\caption{$\rho = 0.20$}
    \end{subfigure}
\begin{subfigure}[b]{0.49\textwidth}
        \includegraphics[width=\textwidth]{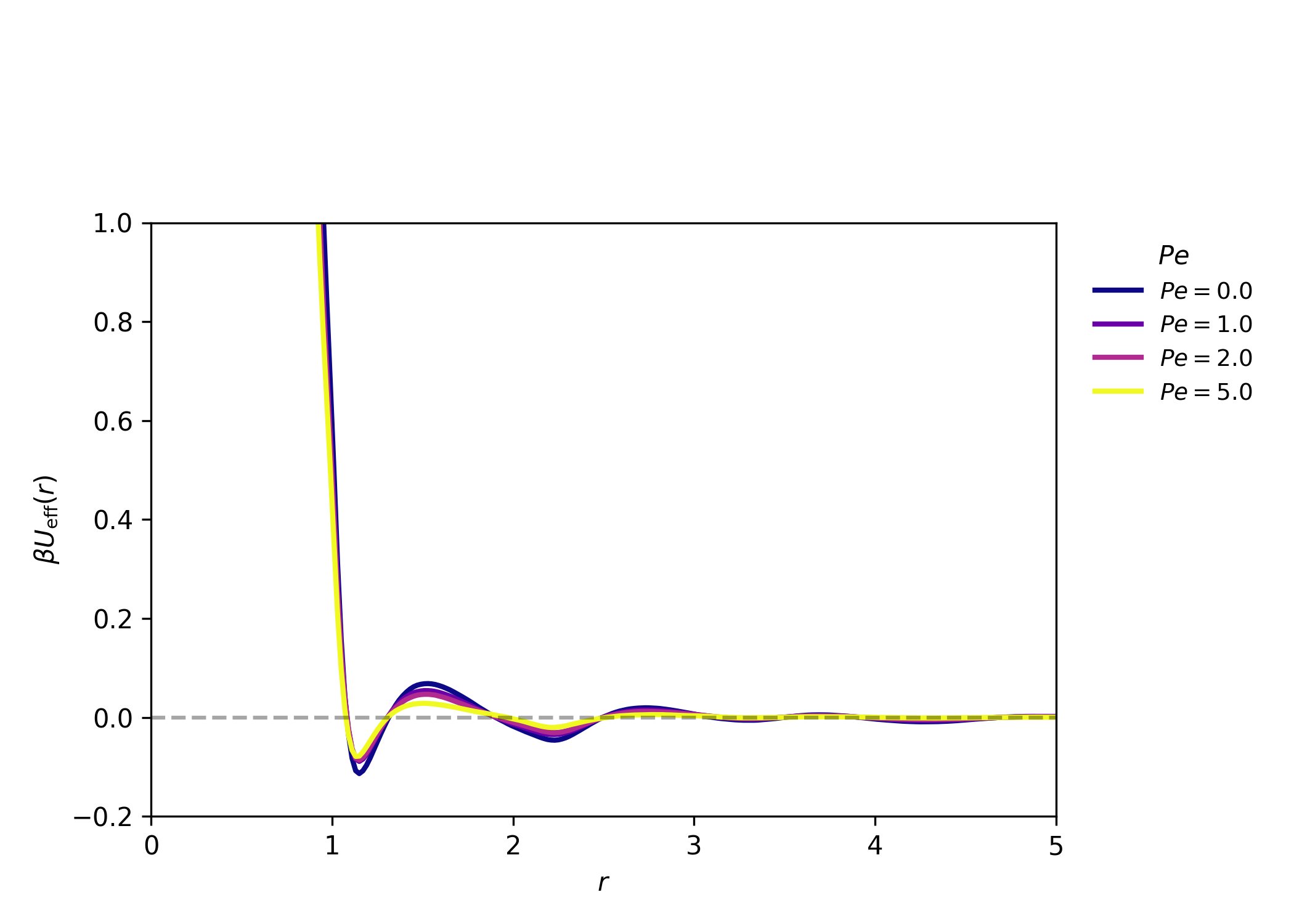}
\caption{$\rho = 0.50$}
    \end{subfigure}

\caption{Effective pair interactions, $\beta u_{\mathrm{eff}}(r)$, at $T=0.20$: (a) passive system for different densities and active systems at (b) $\rho=0.01$, (c) $\rho=0.20$, and (d) $\rho=0.50$ for increasing activities.}
\label{uT020}
    \end{figure}
    
To further clarify the microscopic origin of these structural and dynamical changes, we analyze the effective pair interaction, $\beta u_{\mathrm{eff}}(r)$, obtained using the inverse method proposed in Ref.~\cite{ReesZimmerman2026}. In this approach, the effective potential is defined as the interaction that reproduces the steady-state radial distribution function of the active system, thus incorporating both direct interactions and activity-induced contributions.

For the passive case [Fig.~\ref{uT020}(a)], the effective interaction shows a strong dependence on density. At low density, the effective interaction is largely dominated by the outer repulsive scale created by the Gaussian term. At intermediate densities, it exhibits the typical shoulder–ramp shape, while at higher densities it develops a short-range attraction and long-range repulsion (SALR)-like form. The oscillations at larger distances reflect the structural ordering observed in the RDF. We focus here on three representative densities: $\rho = 0.01$ (gas-like regime), $\rho = 0.20$ (maximum anomaly), and $\rho = 0.50$ (dense fluid).

At $\rho = 0.01$ [Fig.~\ref{uT020}(b)], the passive effective potential is essentially governed by the Gaussian contribution, resulting in an interaction in which the repulsive shoulder is barely developed and particles preferentially occupy the outer scale. As activity increases, $\beta u_{\mathrm{eff}}(r)$ gradually recovers the core-softened profile, with the appearance of a ramp and a shoulder. At high activity ($Pe=5.0$), the effective interaction recovers the characteristic ramp-shoulder shape of the original core-softened potential, although with a reduced shoulder height of approximately $\beta u_{\mathrm{eff}}\simeq0.50$. This indicates that activity does not eliminate the two-length-scale nature of the interaction. Instead, it lowers the effective energetic accessibility associated with occupying the inner length scale.

The effective potential further reveals that increasing activity has an effect analogous to increasing density in equilibrium core-softened systems. As $Pe$ increases, the population of particles gradually shifts from configurations associated with the second length scale toward configurations corresponding to the first length scale. In this sense, activity acts as an effective compression mechanism, promoting local environments that would otherwise become favorable only at higher densities.

At $\rho = 0.20$ [Fig.~\ref{uT020}(c)], the passive system exhibits a core-softened interaction with a repulsive shoulder located at approximately $\beta u_{\mathrm{eff}}\simeq0.80$. Since the energetic cost of crossing the shoulder is comparable to the thermal energy, both local environments remain thermally accessible and populated, as we have shown in Fig.\ref{rdfT020}(c), with a small occupation at the first length scale giving rise to competition between local structures. As the activity increases, the overall ramp-shoulder shape is preserved, but the height of the shoulder decreases markedly, reaching approximately $\beta u_{\mathrm{eff}}\simeq0.20$ at $Pe=5.0$. Consequently, the energetic penalty associated with occupying the inner length scale becomes smaller than the thermal energy, facilitating transitions from the outer to the inner scale. Therefore, activity lowers the energy barrier separating the two local environments and promotes a gradual population transfer toward the first length scale. In this sense, self-propulsion acts as an effective compression mechanism, driving the system along a structural pathway analogous to that produced by increasing density in equilibrium core-softened fluids.

For $\rho = 0.50$ [Fig.~\ref{uT020}(d)], the effective interaction has largely lost the pronounced ramp-shoulder character observed at lower densities. At this density, packing effects have largely transferred the particle population from the outer to the inner length scale, so that the local structure is already dominated by nearest-neighbor configurations associated with the first scale. Consequently, the competition between the two characteristic distances, which is responsible for the anomalous behavior at lower densities, is already strongly diminished in the passive system.

Increasing activity therefore produces only minor modifications in the short-range part of the interaction. Instead, its primary effect is to reduce the oscillations of $\beta u_{\mathrm{eff}}(r)$ at larger distances, indicating a weakening of structural correlations beyond the first coordination shell. In this high-density regime, self-propulsion no longer acts by promoting a transfer of populations between the two scales, since this structural reorganization has already been accomplished by compression. Rather, activity mainly reduces the extent of medium- and long-range ordering while leaving the local environment largely unchanged. Then, the effective interactions consistently show that self-propulsion preserves the two-length-scale character of the interaction while progressively reducing the effective distinction between the competing local environments. The resulting population transfer toward the inner length scale explains the gradual suppression of the anomalous structural and dynamical response discussed above.

\subsection{Non-Anomalous Region - $T = 1.00$}

\begin{figure}[h]
\centering
\begin{subfigure}[b]{0.49\textwidth}
        \includegraphics[width=\textwidth]{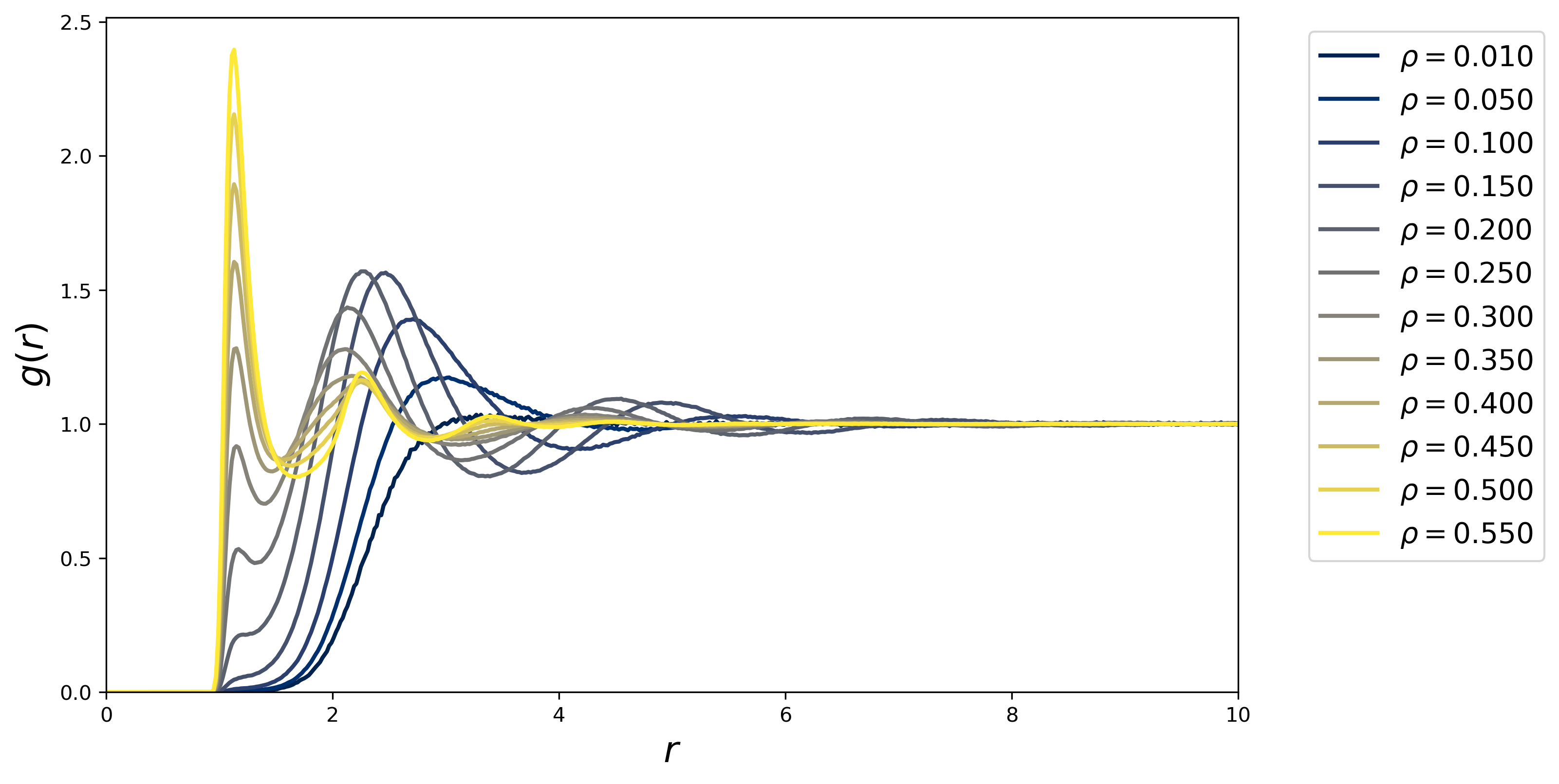}
\caption{}
    \end{subfigure}
\begin{subfigure}[b]{0.49\textwidth}
        \includegraphics[width=\textwidth]{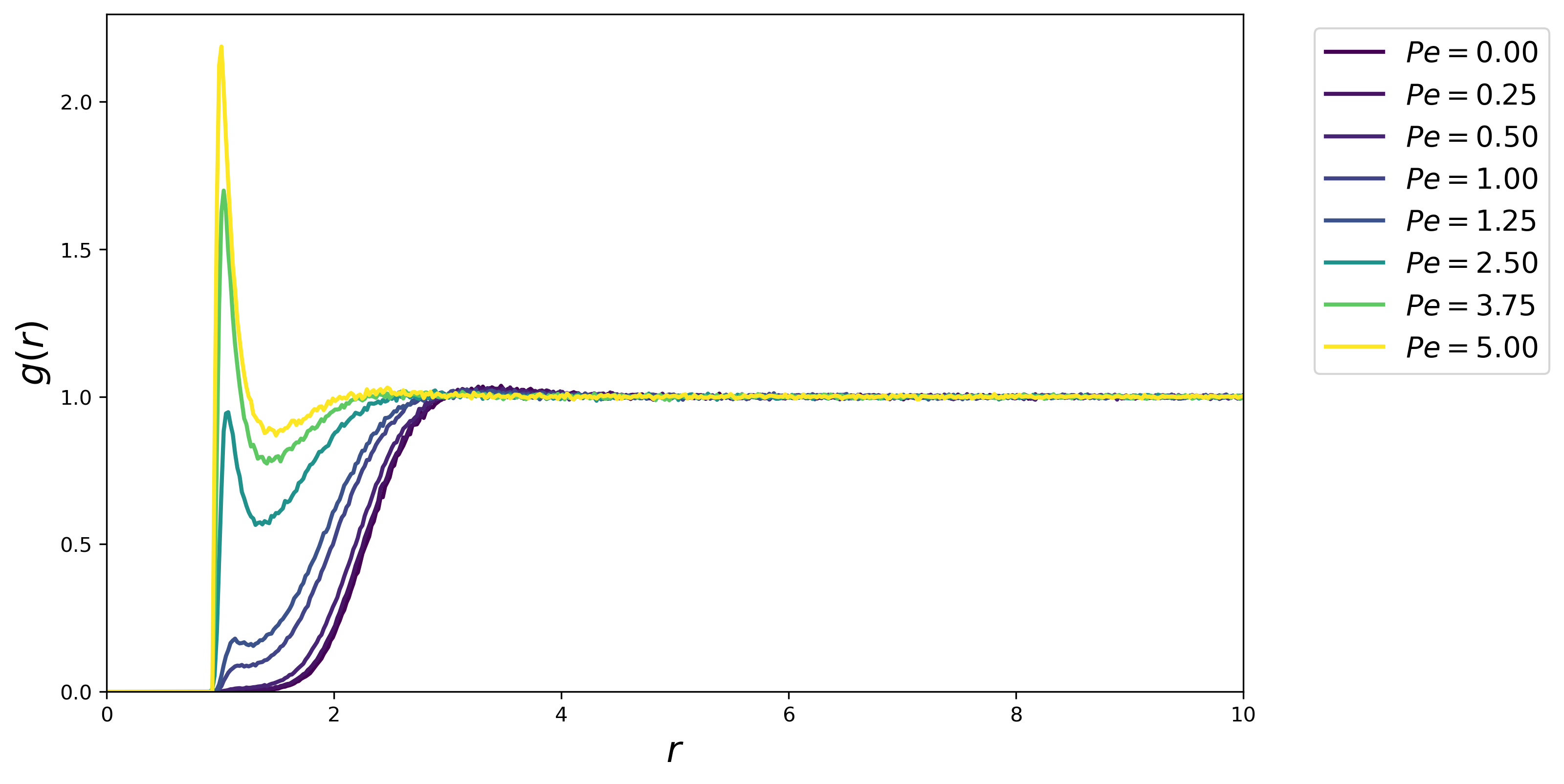}
\caption{}
    \end{subfigure}
    \begin{subfigure}[b]{0.49\textwidth}
        \includegraphics[width=\textwidth]{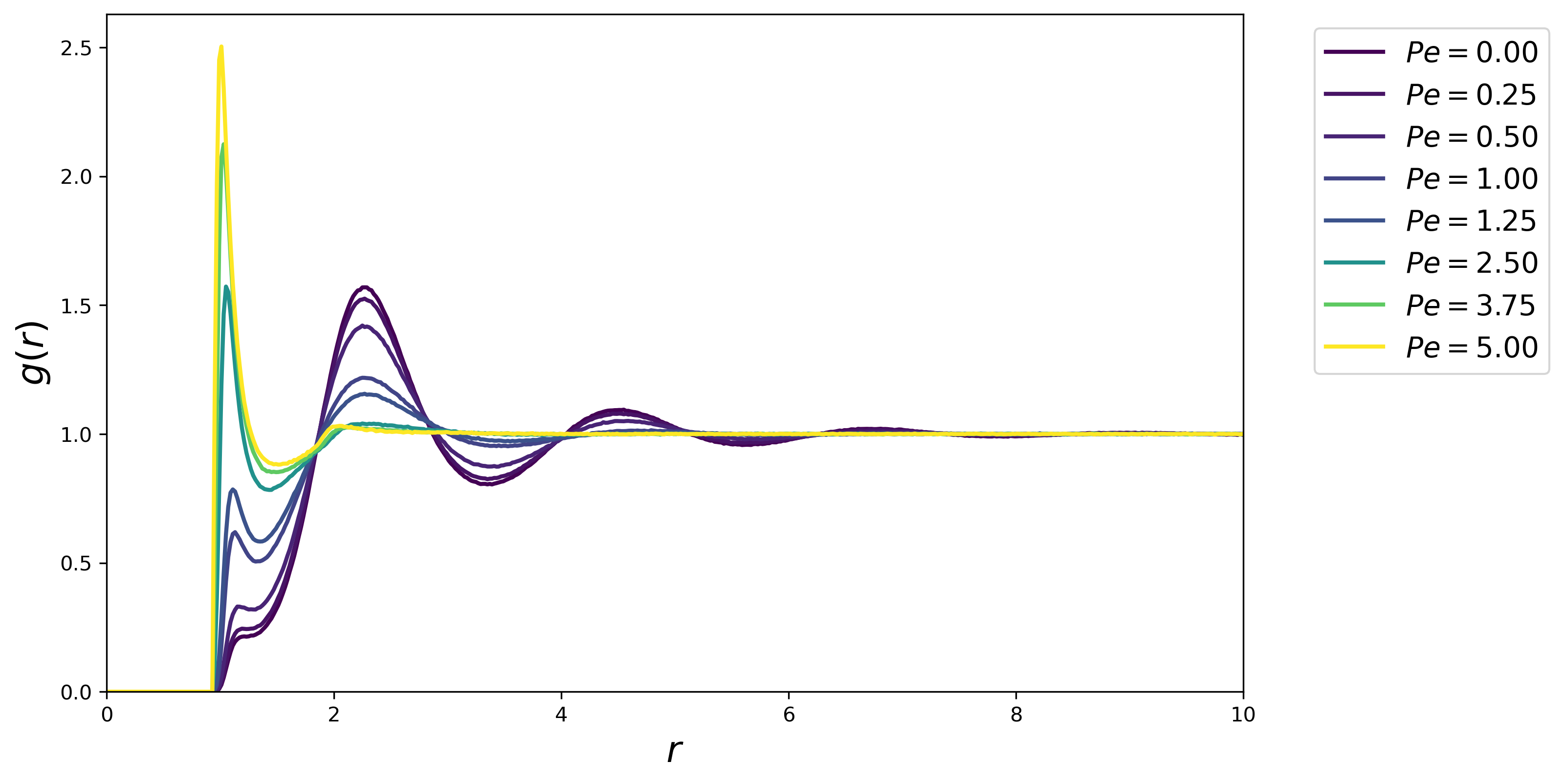}
\caption{}
    \end{subfigure}
    \begin{subfigure}[b]{0.49\textwidth}
        \includegraphics[width=\textwidth]{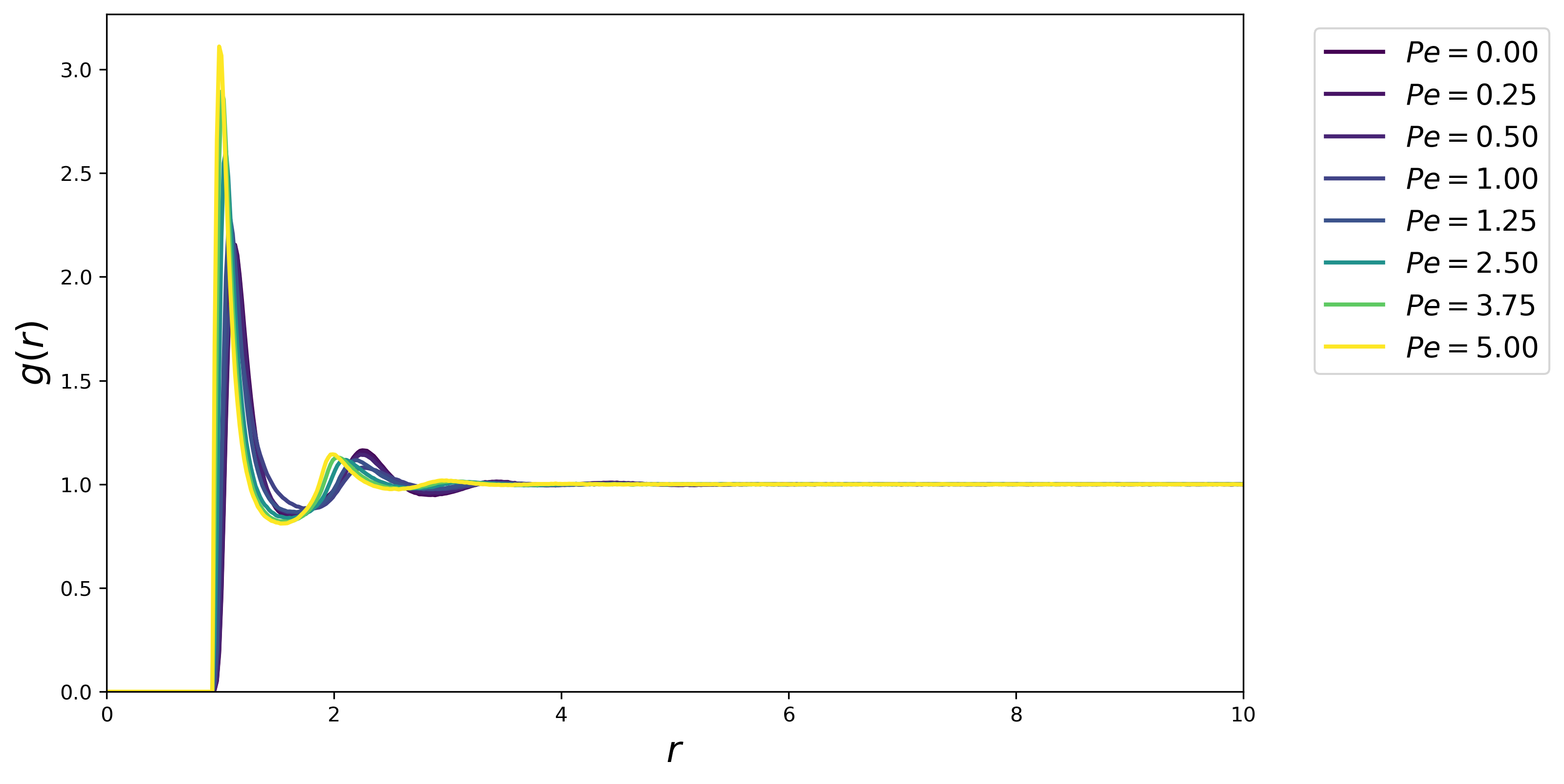}
\caption{}
    \end{subfigure}
\caption{(a) Radial distribution functions at $T=1.00$ for the passive system and different densities. Radial distribution functions for increasing activity at (b) $\rho=0.01$, (c) $\rho=0.20$, and (d) $\rho=0.50$.}
\label{rdfT100}
\end{figure}
    
We now analyze the system at $T=1.00$, where the passive fluid no longer exhibits thermodynamic or dynamic anomalies. At this temperature, thermal fluctuations are sufficiently strong to promote continuous rearrangements between the inner and outer length scales. As a result, the energetic distinction between the two local environments is reduced, and the competition responsible for anomalous behavior becomes weaker.

For the passive case [Fig.~\ref{rdfT100}(a)], the RDF shows a small but finite population at the first length scale. Unlike the low-temperature regime, thermal fluctuations at $T=1.00$ are sufficient to populate the inner scale even at low densities, indicating that the distinction between the two characteristic distances is already reduced. As density increases, the first peak becomes more pronounced, reflecting stronger short-range correlations.

At $\rho = 0.01$ [Fig.~\ref{rdfT100}(b)], activity further increases the occupation of the first length scale. Although the system remains weakly structured, self-propulsion favors shorter interparticle distances, reinforcing the effect of thermal fluctuations.

At $\rho = 0.20$ [Fig.~\ref{rdfT100}(c)], the impact of activity is more evident. The first peak grows significantly, while the second peak is strongly reduced and the third peak nearly disappears, indicating a loss of medium-range organization. Interestingly, this density remains the one most sensitive to activity, corresponding to the same density range ($\rho\approx0.20$--$0.30$) that exhibits anomalous behavior at low temperature and a structural crossover at high temperature. 

At $\rho = 0.50$ [Fig.~\ref{rdfT100}(d)], activity again enhances the occupation of the first length scale. In this regime, however, the second peak does not decrease in height but shifts to larger distances as $Pe$ increases, indicating a reorganization of local packing rather than a suppression of the outer scale. This suggests that, at high density, activity mainly affects correlations beyond the first neighbors, while short-range structure remains dominated by excluded-volume effects.

\begin{figure}[h]
\centering
\begin{subfigure}[b]{0.49\textwidth}
        \includegraphics[width=\textwidth]{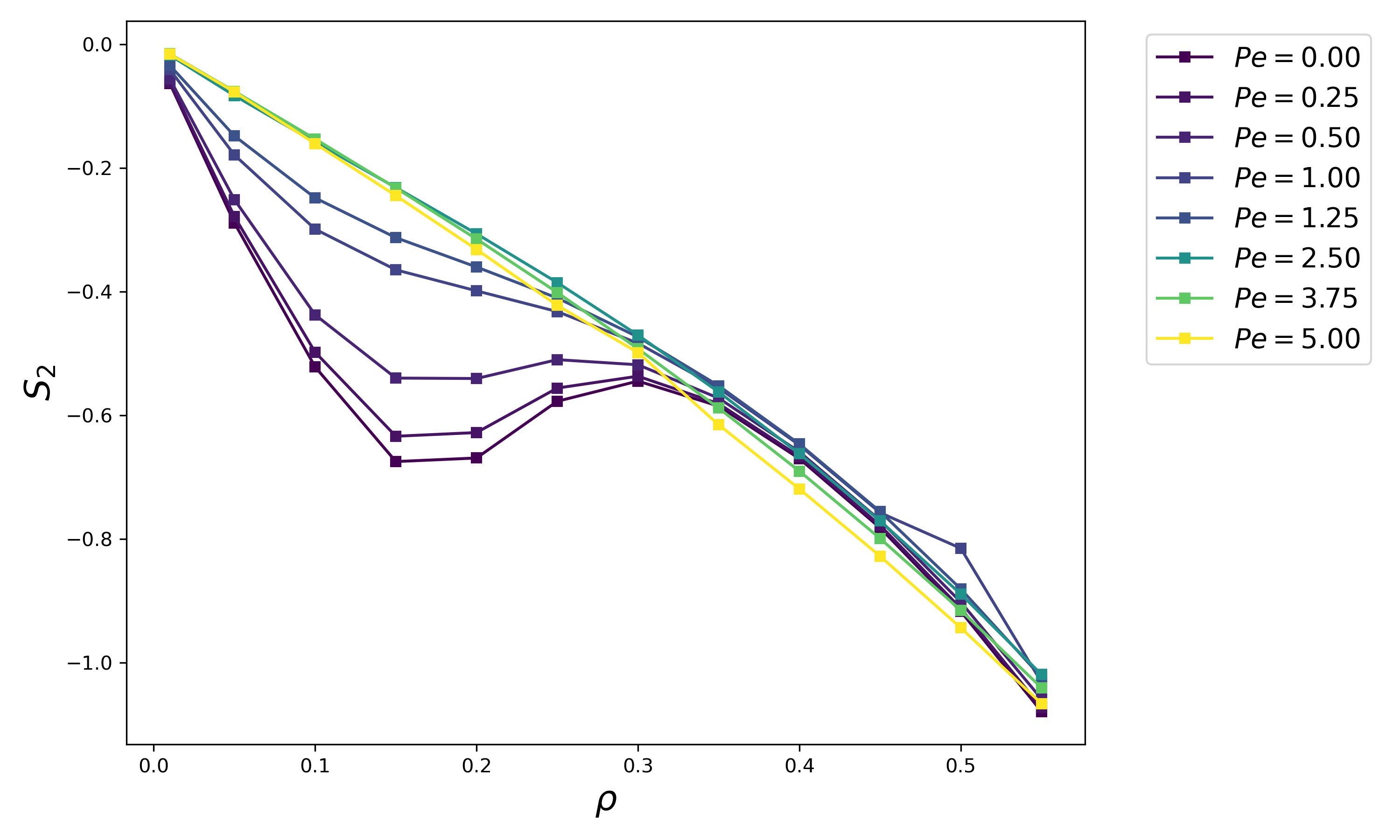}
\caption{}
    \end{subfigure}
\begin{subfigure}[b]{0.49\textwidth}
        \includegraphics[width=\textwidth]{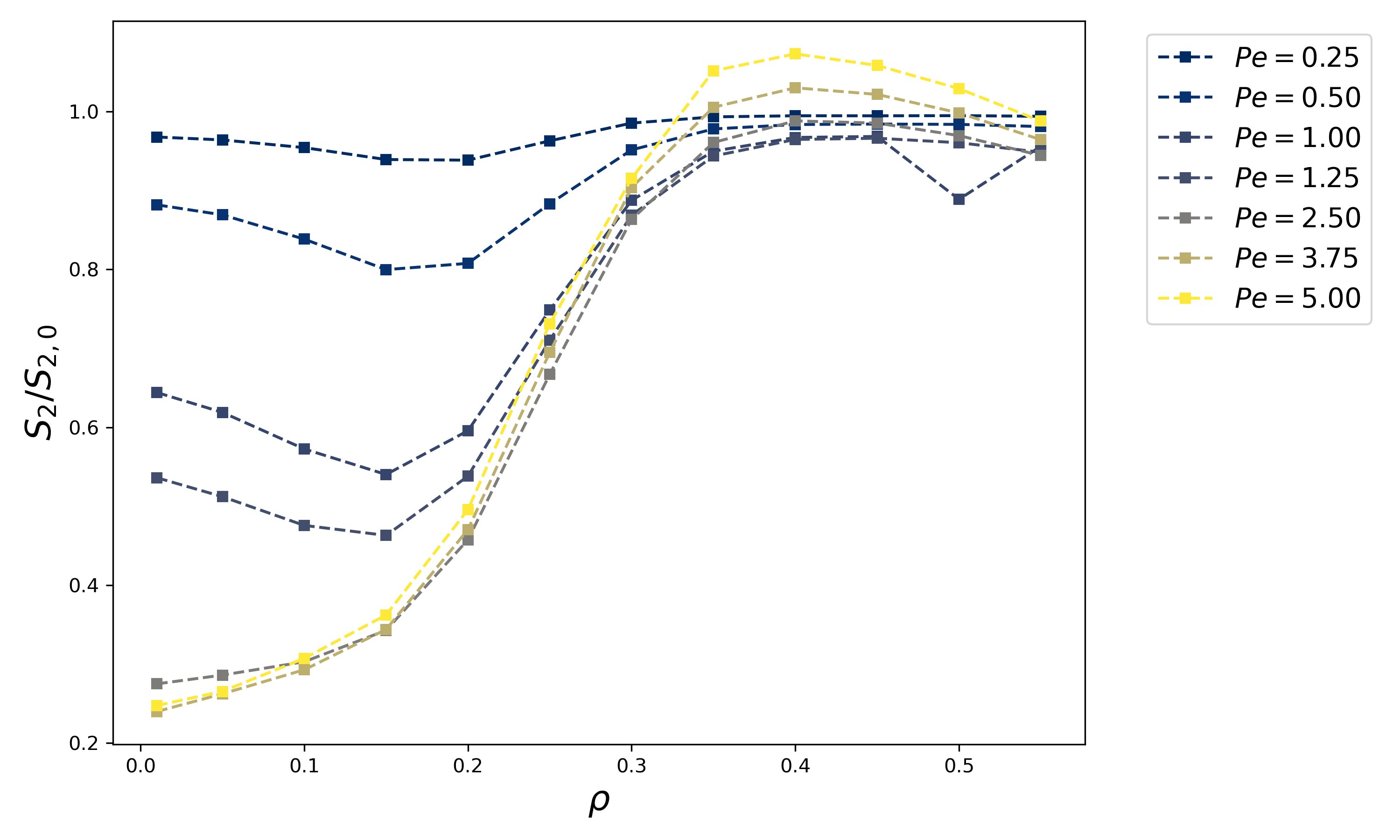}
\caption{}
    \end{subfigure}
\caption{(a) Pair excess entropy $S_2$ as a function of density for different activities at $T=1.00$. (b) Pair excess entropy normalized by its passive counterpart, $S_2/S_{2,0}$.}
\label{s2T100}
    \end{figure}

Figure~\ref{s2T100} shows the pair excess entropy $S_2$ at $T=1.00$. For the passive fluid [Fig.~\ref{s2T100}(a)], no clear structural anomaly is observed, as $S_2$ no longer exhibits the minimum--maximum pattern found at low temperature. Instead, the curve displays a change in curvature around $\rho\approx0.20$--$0.30$. This behavior indicates that the two characteristic distances remain present, but the energetic distinction between them has become sufficiently small that particles can continuously rearrange between the two local environments. Consequently, the sharp structural reorganization observed at $T=0.20$ is replaced by a smooth crossover.

This interpretation is also reflected in the density dependence of $S_2$. Up to $\rho\approx0.30$, the entropy decreases slowly because both length scales remain accessible and particle rearrangements between them are still possible. At higher densities, the decay becomes nearly linear, indicating that the local structure is increasingly dominated by configurations associated with the first length scale and by excluded-volume effects.

Increasing activity further smooths this crossover. The $S_2(\rho)$ curves become progressively more linear, suggesting that self-propulsion enhances the redistribution of particles toward the inner scale and further reduces the remaining structural distinction between the two local environments.

The normalized quantity $S_2/S_{2,0}$ [Fig.~\ref{s2T100}(b)] nevertheless retains a non-monotonic dependence on density. The largest deviations occur around $\rho\approx0.20$--$0.30$, corresponding to the density range that displays anomalous behavior at low temperature and a structural crossover at $T=1.00$. Therefore, even in the absence of anomalies in the absolute quantities, this density interval remains the one most susceptible to activity-induced structural changes.

\begin{figure}[h]
\centering
\begin{subfigure}[b]{0.49\textwidth}
        \includegraphics[width=\textwidth]{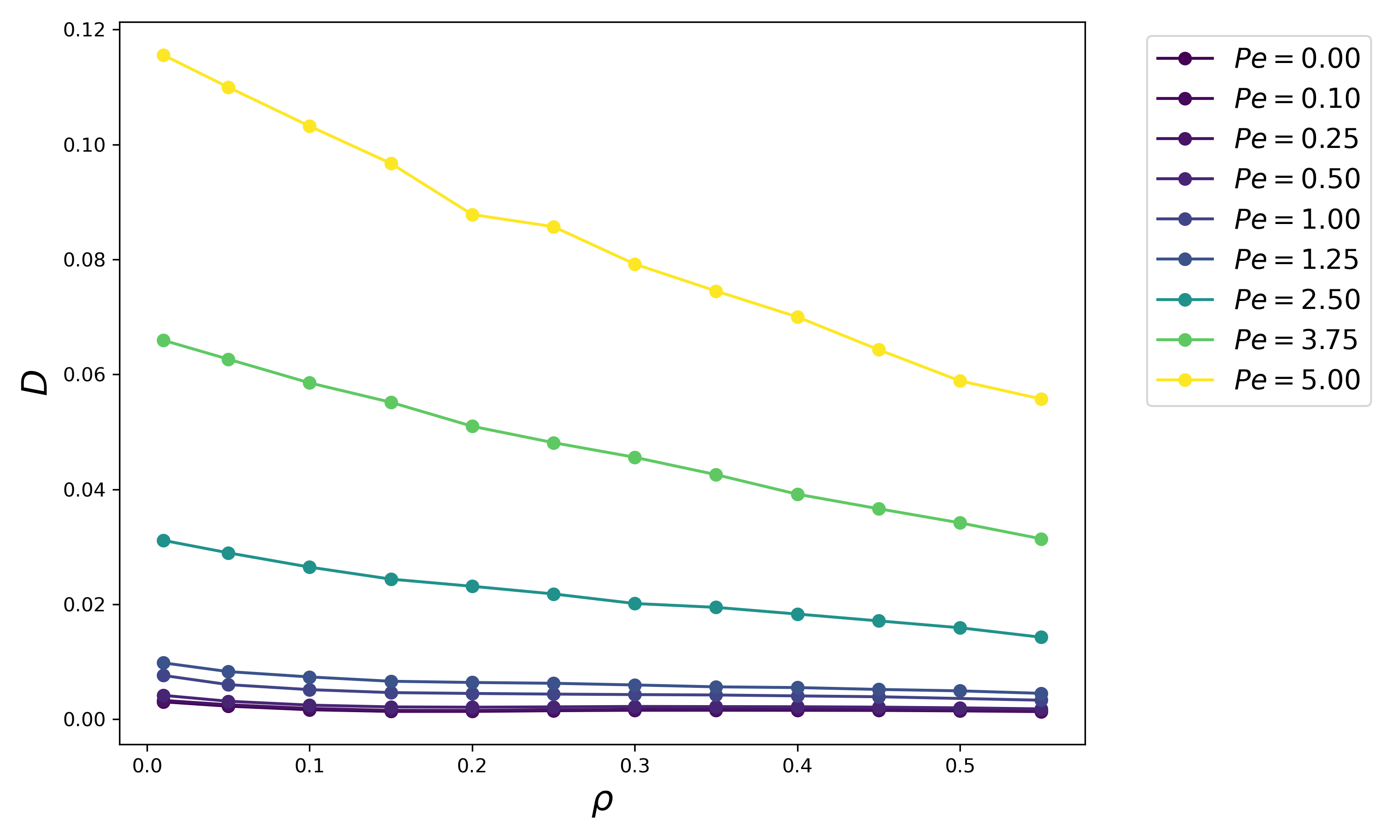}
\caption{}
    \end{subfigure}
\begin{subfigure}[b]{0.49\textwidth}
        \includegraphics[width=\textwidth]{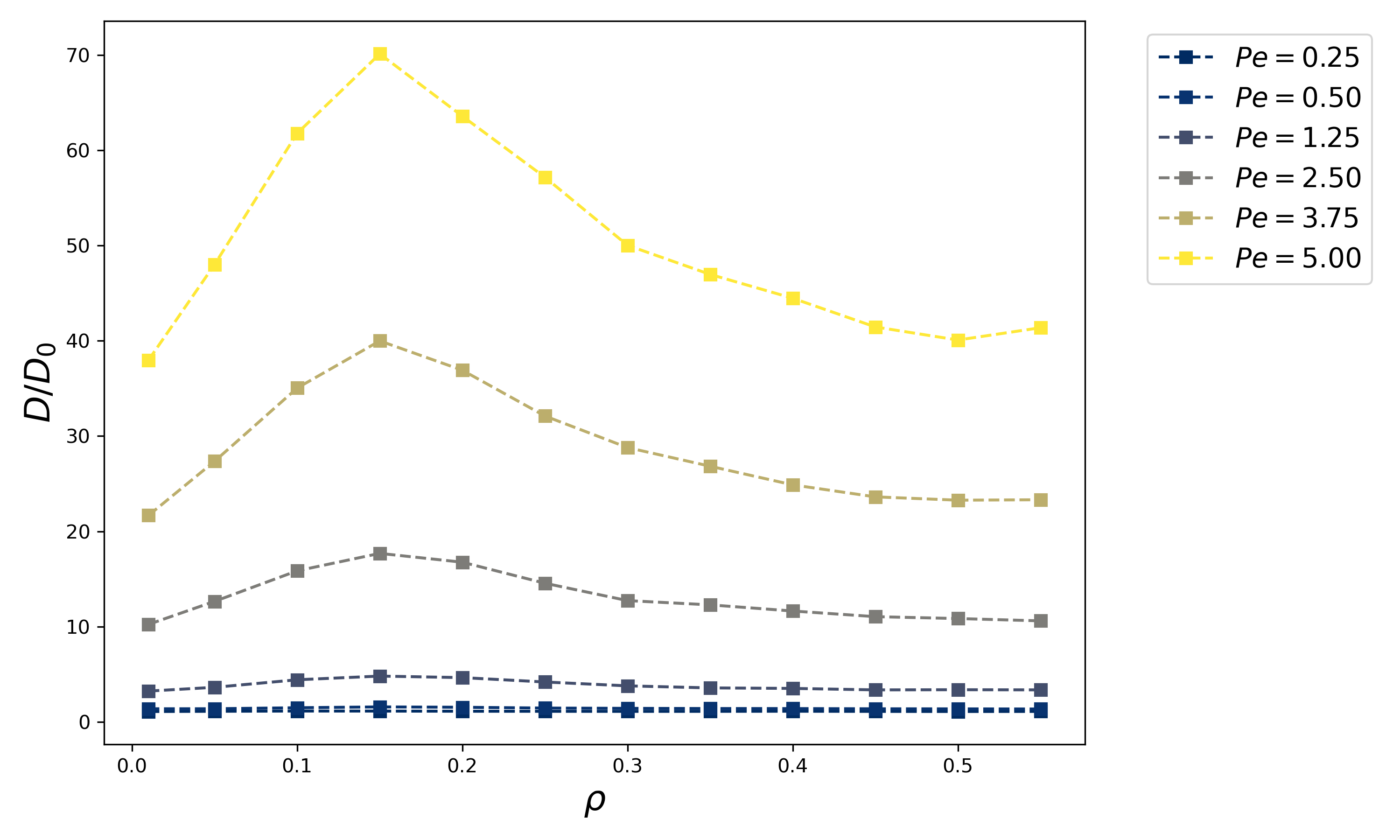}
\caption{}
    \end{subfigure}
    \caption{(a) Diffusion coefficient $D$ at $T = 1.00$ as function of density $\rho$ for distinct activities. (b) $D$ normalized by the passive diffusion, $D_{0}$, as function of $\rho$. }
   \label{DT100}
    \end{figure}

A similar picture emerges from the diffusion coefficient shown in Fig.~\ref{DT100}. For the passive fluid [Fig.~\ref{DT100}(a)], $D$ decreases monotonically with density but still exhibits a slight change in curvature around $\rho\approx0.20$--$0.30$, mirroring the behavior of $S_2$. As activity increases, this feature gradually disappears and the dependence becomes fully monotonic.

The normalized diffusion coefficient, $D/D_0$ [Fig.~\ref{DT100}(b)], further emphasizes the special role of this density range. A maximum is observed around $\rho\approx0.20$, indicating that self-propulsion produces the largest relative enhancement of particle mobility precisely where the two characteristic distances remain most distinguishable. The coincidence between the extrema of $S_2/S_{2,0}$ and the maximum of $D/D_0$ shows that, even though thermal fluctuations have already weakened the energetic competition between the two local environments, the density range around $\rho\approx0.20$--$0.30$ remains the region where activity most efficiently promotes structural rearrangements between the two characteristic scales.

\begin{figure}[h]
\centering
    \begin{subfigure}[b]{0.49\textwidth}
        \includegraphics[width=\textwidth]{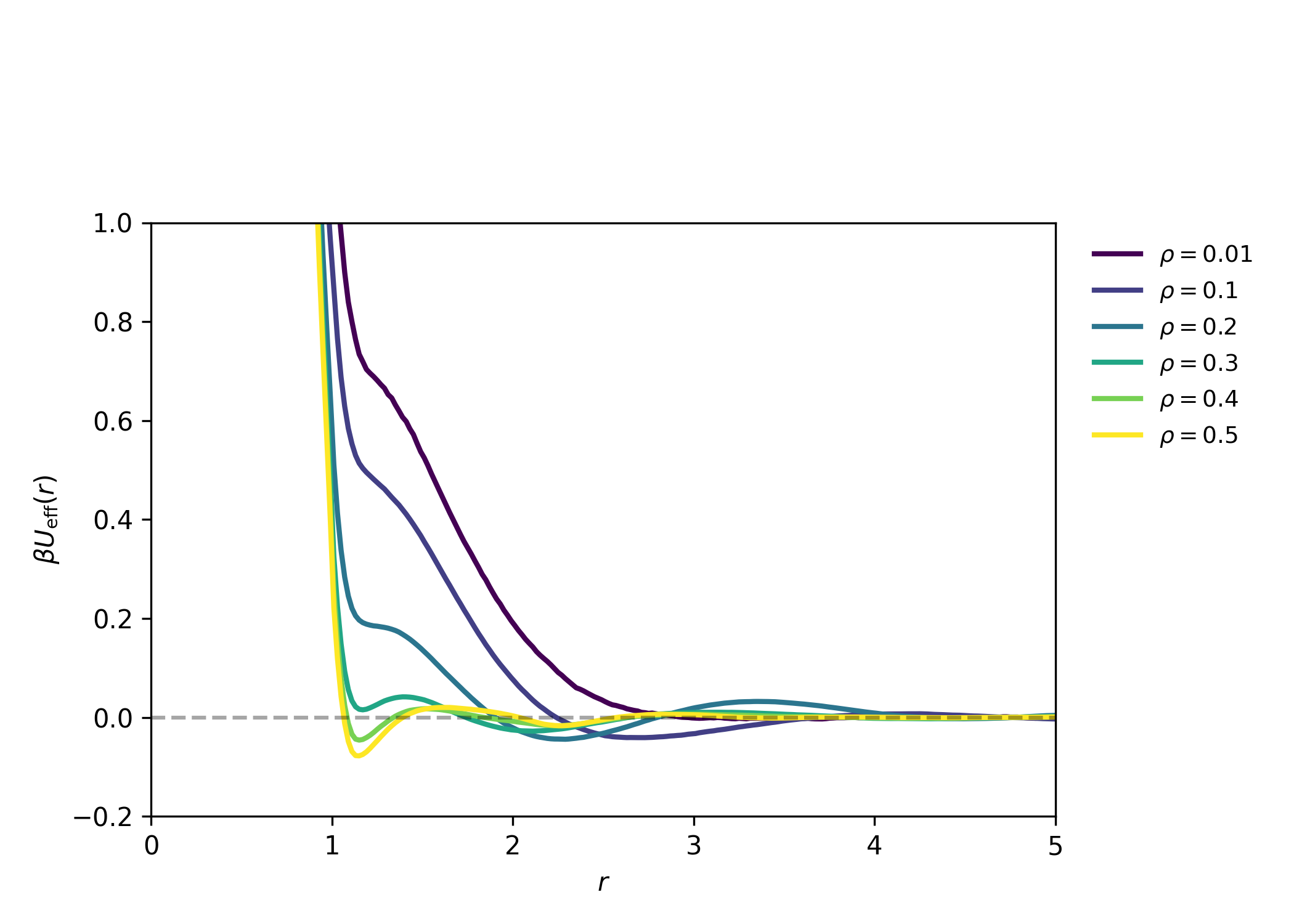}
\caption{$Pe = 0.0$}
    \end{subfigure}
\begin{subfigure}[b]{0.49\textwidth}
        \includegraphics[width=\textwidth]{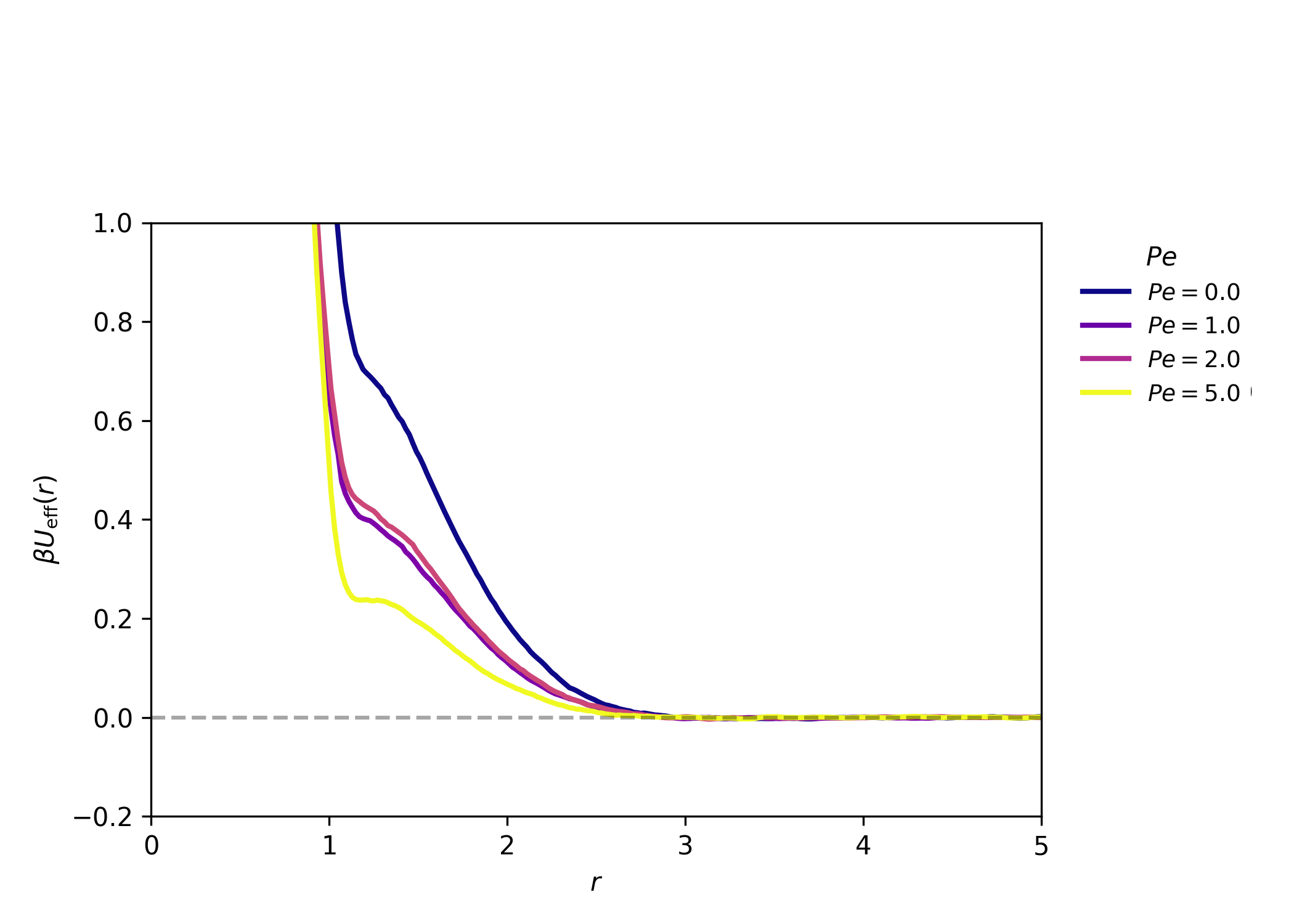}
\caption{$\rho = 0.01$}
    \end{subfigure}
\begin{subfigure}[b]{0.49\textwidth}
        \includegraphics[width=\textwidth]{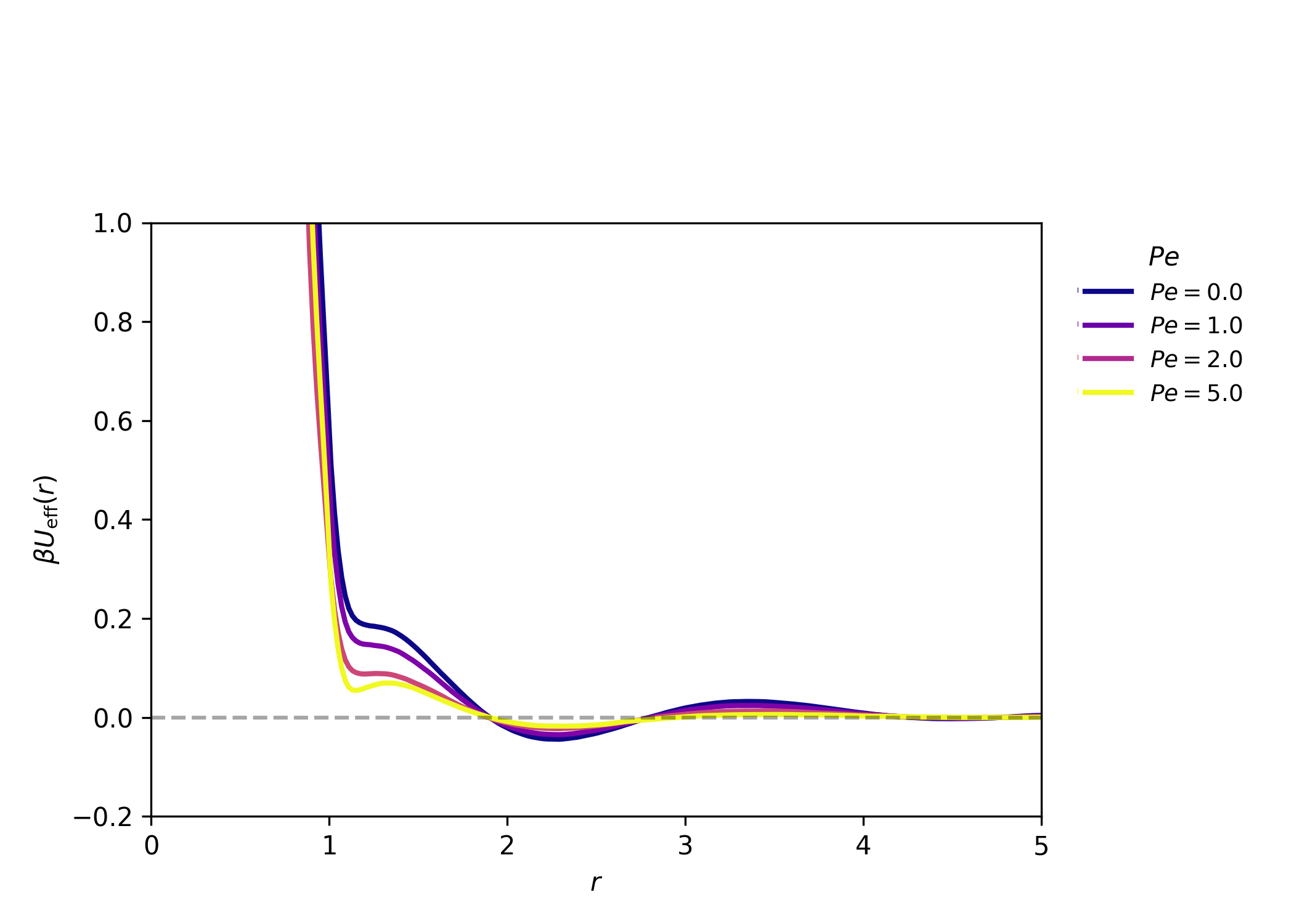}
\caption{$\rho = 0.20$}
    \end{subfigure}
\begin{subfigure}[b]{0.49\textwidth}
        \includegraphics[width=\textwidth]{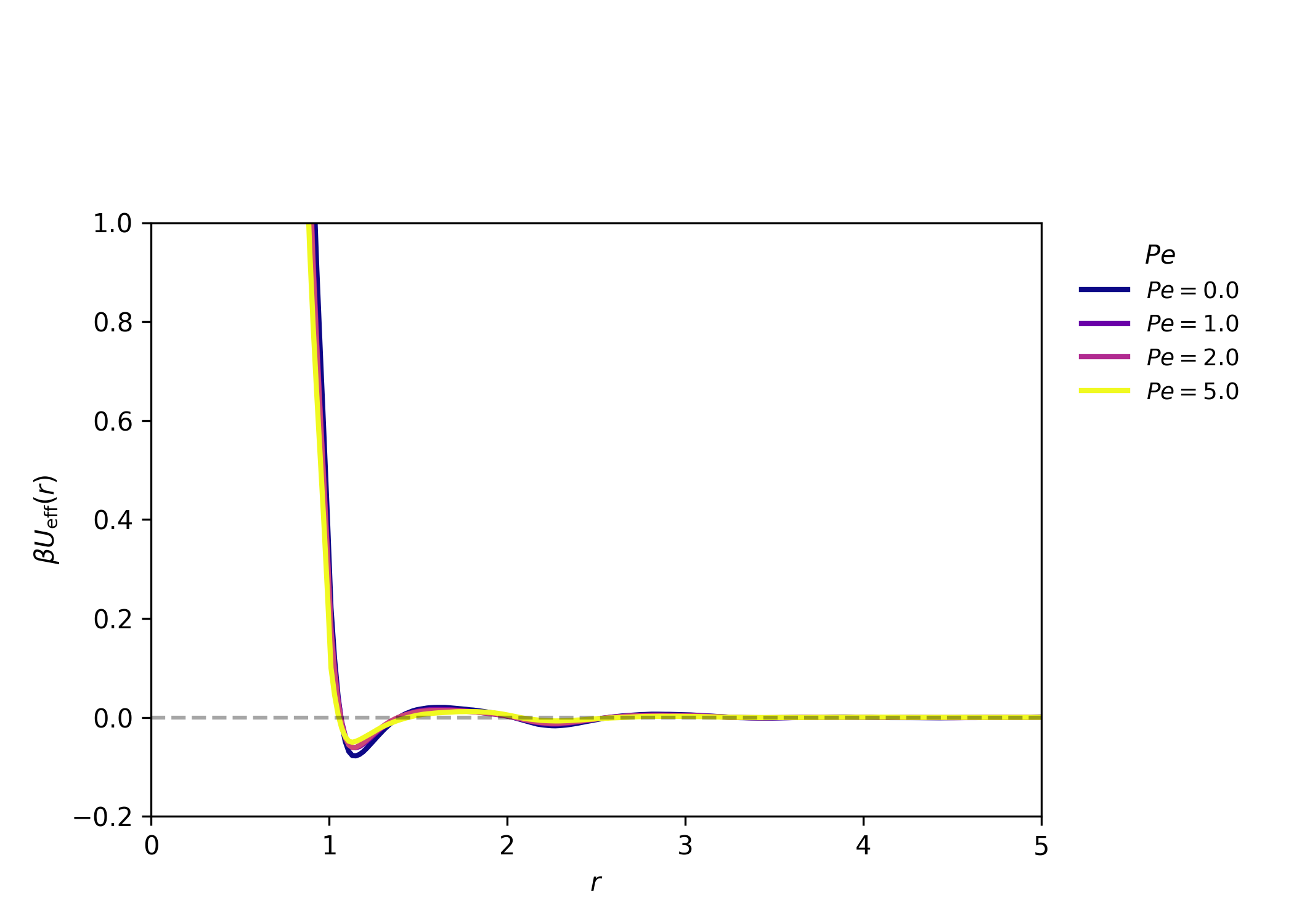}
\caption{$\rho = 0.50$}
    \end{subfigure}

\caption{Effective pair interactions, $\beta u_{\mathrm{eff}}(r)$, at $T=1.00$: (a) passive system for different densities and active systems at (b) $\rho=0.01$, (c) $\rho=0.20$, and (d) $\rho=0.50$ for increasing activities.}
\label{uT100}
    \end{figure}
  
The effective pair interactions at high temperature are shown in Fig.~\ref{uT100}. In the passive case [Fig.~\ref{uT100}(a)], the interaction retains the characteristic ramp-shoulder shape of the core-softened potential, although the repulsive shoulder is lower than at $T=0.20$, reaching approximately $\beta u_{\mathrm{eff}}\simeq0.75$ at low density. As the density increases, the height of the shoulder decreases continuously and eventually disappears at high density, indicating that thermal fluctuations and compression favor configurations associated with the inner length scale. Consequently, the competition between the two characteristic distances is already weaker than at low temperature.

At $\rho = 0.01$ [Fig.~\ref{uT100}(b)], the passive effective interaction exhibits a well-defined but relatively shallow shoulder. Increasing activity preserves the ramp-shoulder form of the interaction while progressively reducing the height of the shoulder, which reaches approximately $\beta u_{\mathrm{eff}}\simeq0.20$ at $Pe=5.0$. Then, it further lowers the effective energetic cost associated with occupying the inner length scale, promoting a redistribution of populations toward shorter distances. 

At $\rho = 0.20$ [Fig.~\ref{uT100}(c)], the passive interaction still displays a weak ramp-shoulder structure, with a shoulder height of approximately $\beta u_{\mathrm{eff}}\simeq0.20$. Consequently, the energetic penalty associated with crossing from the outer to the inner length scale is already smaller than the thermal energy. Increasing activity further lowers the shoulder to values below $\beta u_{\mathrm{eff}}\simeq0.10$ at $Pe=5.0$, making transitions between the two local environments occur with smaller enthalpic cost. Simultaneously, the oscillations of $u_{\mathrm{eff}}(r)$ at larger distances become weaker, indicating a reduction of medium- and long-range correlations. Therefore, activity acts primarily by enhancing the effective compression already induced by thermal fluctuations, promoting local structural rearrangements while reducing spatial correlations beyond the first coordination shell.

At $\rho = 0.50$ [Fig.~\ref{uT100}(d)], the interaction no longer exhibits a pronounced ramp-shoulder character and is already dominated by configurations associated with the first length scale. Consequently, activity produces only minor modifications in the short-range part of the interaction. As in the low-temperature case, its primary effect is to suppress oscillations at larger distances, indicating weaker medium- and long-range correlations while leaving the local environment essentially unchanged.

The effective interactions help rationalize the absence of thermodynamic and dynamic anomalies at $T=1.00$. At this temperature, the repulsive shoulder is already sufficiently low that transitions between the two characteristic distances occur with little energetic cost. Thermal fluctuations alone are therefore able to continuously redistribute particles between the inner and outer scales. Consequently, the coexistence of two competing local environments, which underlies the anomalies at low temperature, becomes much weaker. Self-propulsion further lowers the energetic barrier and reinforces this tendency, but it does not introduce a qualitatively new mechanism. Instead, temperature and activity act cooperatively, both favoring configurations associated with the first characteristic distance.

\section{Conclusion}
In this work, we investigated how activity modifies a fluid with competing interaction length scales, modeled by a ramp-like core-softened potential. Our results provide a microscopic picture of how nonequilibrium driving affects the structural and dynamical mechanisms associated with water-like anomalies.

At low temperature, the anomalies originate from the competition between two characteristic distances. The effective interactions reveal that self-propulsion lowers the energetic barrier separating these local environments and promotes a redistribution of particle populations toward the inner length scale. Consequently, the range of state points over which both scales remain significantly populated becomes narrower, leading to the progressive suppression of structural and dynamical anomalies. In this regime, activity acts as an effective compression mechanism, driving the system along a structural pathway that closely resembles equilibrium compression.

At higher temperature, thermal fluctuations have already reduced the energetic distinction between the two scales, and thermodynamic and dynamic anomalies are no longer observed in absolute quantities. Nevertheless, both structure and dynamics remain most sensitive to activity in the same density range, $\rho\approx0.20$--$0.30$. The non-monotonic behavior of the normalized quantities $S_2/S_{2,0}$ and $D/D_0$ indicates that the underlying two-scale mechanism remains active, even though its macroscopic signatures become much weaker. In this regime, self-propulsion and thermal fluctuations act cooperatively, both facilitating structural rearrangements toward the inner scale.

Both temperature and activity facilitate transitions between the competing local environments by reducing the effective distinction between the two characteristic length scales. While self-propulsion does not eliminate the two-length-scale nature of the interaction, it modifies the balance between the competing local structures. Interesting, even relatively weak activity is sufficient to substantially suppress the anomalous response, demonstrating that persistent self-propulsion provides an efficient nonequilibrium route to weaken the structural competition responsible for water-like anomalies.

\begin{acknowledgments}

Without public funding this research would be impossible. DFKS acknowledges support from Coordination for the Improvement of Higher Education Personnel (CAPES), finance Code 001. J.R.B. acknowledges financial support from Brazilian National Council for Scientific and Technological Development (CNPq), grant numbers 405479/2023-9, 441728/2023-5, and 304958/2022-0A, and from the CAPES and the Alexander von Humboldt Foundation for financial support through a research fellowship. 
\end{acknowledgments}

\section*{Data Availability Statement}

The simulation input files, analysis scripts, and plotting routines are publicly available through a GitHub repository archived in Zenodo: {\url {https://doi.org/10.5281/zenodo.21068695}}

\bibliography{aipsamp}

\end{document}